\documentclass[11 pt]{article}
\RequirePackage[colorlinks,citecolor=blue,urlcolor=blue]{hyperref}
\usepackage{amssymb}
\usepackage[all,knot,arc,import,poly]{xy}
\usepackage{graphicx}
\usepackage{amsmath}
\usepackage{color}
\usepackage{enumerate}
\usepackage{tikz}
\usetikzlibrary{positioning,shapes,arrows, snakes,fit}
\usetikzlibrary{matrix}
\usepackage{graphicx}
\usepackage{subfig}
\usepackage{moreverb}
\usepackage{hyperref}

\newcommand{\beginsupplement}{%
        \setcounter{table}{0}
        \renewcommand{\thetable}{S\arabic{table}}%
        \setcounter{figure}{0}
        \renewcommand{\thefigure}{S\arabic{figure}}%
     }

\begin{document}
\begin{center}{
\textbf{\Large Secondary Phenotype Analysis in Ascertained Family Designs: Application to the Leiden Longevity Study}\\
}
\end{center}

Renaud TISSIER$^{1}$ , Roula TSONAKA$^{1}$, Simon P. MOOIJAART$^{2}$, Eline SLAGBOOM$^{3}$, Jeanine J. HOUWING-DUISTERMAAT$^{1,4}$
\bigskip

\noindent
$^{1}$Department of Medical Statistics and Bioinformatics, Leiden University Medical Centre, Leiden, The Netherlands.

\noindent
$^{2}$Department of Gerontology and Geriatrics, Leiden University Medical Centre, Leiden, The Netherlands.

\noindent
$^{3}$Department of Molecular Epidemiology, Leiden University Medical Centre, Leiden, The Netherlands.

\noindent
$^{4}$Department of Statistics, University of Leeds, United Kingdom.

\begin{abstract}
The case-control design is often used to test associations between the case-control status and genetic variants. In addition to this primary phenotype a number of additional traits, known as secondary phenotypes, are routinely recorded and typically associations between genetic factors and these secondary traits are studied too. Analysing secondary phenotypes in case-control studies may lead to  biased genetic effect estimates, especially when the marker tested is associated with the primary phenotype and when the primary and secondary phenotypes tested are correlated. Several methods have been proposed in the literature to overcome the problem but they are limited to case-control studies and not directly applicable to more complex designs, such as the multiple-cases family studies. A proper secondary phenotype analysis, in this case, is complicated by the within families correlations on top of the biased sampling design.
We propose a novel approach to accommodate the ascertainment process while explicitly modelling the familial relationships. Our approach pairs existing methods for mixed-effects models with the retrospective likelihood framework and uses a multivariate probit model to capture the association between the mixed type primary and secondary phenotypes. To examine the efficiency and bias of the estimates we performed simulations under several scenarios for the association between the primary phenotype, secondary phenotype, and genetic markers. We will illustrate the method by analysing the association between triglyceride levels and glucose (secondary phenotypes) and genetic markers from the Leiden Longevity study, a multiple-cases family study that investigates longevity.
\end{abstract}

\section{Introduction}
\label{sec1}

In order to understand biological mechanisms underlying disease and health, epidemiological studies measure genetic markers, classical variables, and novel omics datasets and model the relationship between these variables and the phenotype of interest. Here we consider outcome dependent sampling designs with binary outcome variables. In addition to studying these binary (primary) phenotypes, the classical or omics variables are typically also analysed as outcome variables (secondary phenotypes). For example modelling of associations between these traits and genetic factors, such as single-nucleotide polymorphisms (SNPs) or polygenic risk scores (sumscores based on SNPs)\cite{DudbridgeF2013}. However, an important complication which is often ignored is that a proper analysis of the secondary traits should correct for the sampling mechanism on the primary phenotype (Figure \ref{DAG}). In our motivating case study, the Leiden Longevity study (LLS)\cite{Houwing2009} families with at least two long-lived siblings are recruited. Obviously, these families do not represent a random sample from the population and inferences cannot be generalized to the whole population, unless the sampling mechanism is properly modelled. Several datasets are measured in the offspring of the long-lived siblings, namely lipidomics, glycomics, metabolomics, and imaging. These offspring share a part of their genetic variation with the long-lived parent and therefore are expected to represent a healthy subpopulation while the partners represent the population. As data example we will model the effect of genetic factors on the secondary traits glucose and triglyceride levels in the offspring (cases) and their partners (controls). To be able to extrapolate results to the general population, we need to account for the over sampling of long-lived subjects in the families of the LLS. There are several multiple-case family studies. For human longevity, GEHA (Genetics of Healthy aging)\cite{Skytthe2011} used the same study design as the LLS. Other examples are  Genetics in Familial Thrombosis (GIFT with at least two cases with thrombosis) \cite{Visser2013,Tsonaka2012} and the ongoing study from Leiden Family Lab (famlab: https://www.leidenfamilylab.nl) which recruits families with at least two cases with social anxiety disorder. The novel methods presented in this paper will also be essential for modelling secondary phenotypes in these studies.

In the context of case-control studies Monsees \emph{et~al.}\cite{Monsees2009} showed that bias can occur when estimating the SNP effect on secondary phenotypes if the primary and secondary phenotypes are associated. This is often the case because both outcomes are measured on the same subjects and secondary phenotypes are typically chosen for their potential associations with the primary phenotype. They also showed that the amount of bias is dependent on the prevalence of the primary phenotype, the strength of the association between the primary and secondary phenotypes, and the association between the tested marker and the primary trait (see Figure \ref{DAG}).

To deal with the bias problem, investigators first used ad hoc methods i.e. using controls only, cases only, combined data of cases and controls or joint analysis of cases and controls adjusting for the case-control status. However, several authors showed that these simple approaches can lead to false positive results \cite{Monsees2009,Lee1997,Lin2009}. This is due to the sampling design, namely, the secondary phenotype data are not sampled according to the case-control design as the primary phenotype. Several sophisticated methodologies have been developed to correct for the sampling mechanisms and provide unbiased genetic effect estimates: (i) inverse-probability-of-sampling-weighting approaches \cite{Monsees2009,Richardson2007, Schifano2013} which correct for the sampling mechanism by weighting appropriately individuals in case-control studies, (ii) retrospective likelihood-based approaches which indirectly adjust for ascertainment \cite{Lin2009,He2011}, and (iii) a weighted combination of two estimates obtained with the retrospective likelihood approach in the presence or not of an interaction between SNPs and primary phenotypes \cite{Li2012}.

Even though these approaches can successfully correct for the biased design used to collect the data, they are not directly applicable to more complex designs such as the LLS which motivates this work. In particular, inverse probability weighting approaches require knowledge of the sampling weights for each family. These weights are not available for the LLS because it is unknown what the prevalence of families with at least two nonagenarians is in the population. In addition, the correlations between the family members cannot be ignored and therefore it is evident that statistical methodology for proper secondary phenotypes analysis in this context is needed. To this end, under the retrospective likelihood framework, we develop a multivariate probit regression model inspired by the work of Najita \emph{et~al.}\cite{Najita2009} to model jointly the distribution of the primary and secondary phenotype. This approach allows us to deal with the ascertainment issue while taking into account the individual relatedness and the genetic and environmental variations.

The paper is organised as follows: in Section 2, we present the retrospective likelihood approach to correct for the over sampling of long-lived subjects and the multivariate probit regression model for the joint modelling of the mixed type primary and secondary phenotypes. In Section 3, we evaluate empirically the performance of the method in terms of bias and efficiency and contrast it with the naive approach which ignores the sampling mechanism. Finally, in Section 4 we illustrate the potential of our proposed method in the analysis of triglyceride levels and glucose in the LLS.

\section{Methods}
\label{sec2}

\subsection{Retrospective likelihood approach}

Let $N$ be the total number of families in the study. For the family $i$ ($i = 1\ldots N$) of size $n_{i}$, let $Y_{i}$, $X_{i}$ and $G_{i}$ be the $n_{i}\times1$ vectors for the case-control status, the secondary phenotype and the genotype, respectively.
Motivated by the LLS, we will work under the retrospective likelihood approach to correct for the ascertainment of the families. Such an approach is attractive when modelling the ascertainment mechanism is not straightforward, as in the LLS where sampling depends on the previous generation (an example of a pedigree in LLS is shown in Figure \ref{LLS pedigree}). In fact the retrospective likelihood approach implicitly corrects for the ascertainment mechanism, under the assumption that the ascertainment depends only on the primary phenotype $Y$. In particular, for the $i$th family it holds:
\begin{equation}
P\left(X_{i},G_{i}\mid  Y_{i},Asc\right)=\frac{P\left(Asc\mid  Y_{i},G_{i},X_{i}\right)P\left(G_{i},X_{i}\mid Y_{i}\right)}{P\left(Asc\mid Y_{i}\right)}=P\left(X_{i},G_{i}\mid Y_{i}\right),
\label{Implicit correction for ascertainment}
\end{equation}
with $Asc$ the ascertainment process. By applying Bayes rule we obtain:
\begin{equation}
P\left(X_{i},G_{i}\mid  Y_{i}\right)=\frac{P\left(X_{i},Y_{i}\mid G_{i}\right)P\left(G_{i}\right)}{P\left(Y_{i}\right)}=\frac{P\left(X_{i},Y_{i}\mid G_{i}\right)P\left(G_{i}\right)}{\sum_{g\in G}P\left(Y_{i}\mid g\right)P\left(g\right)}.
\label{Likelihood obtained}
\end{equation}
To fully specify (\ref{Likelihood obtained}) we need to model properly: the marginal joint distribution of the primary phenotype with the secondary phenotype P($X_{i},Y_{i}\mid G_{i}$), the marginal probability of the primary phenotype P($Y_{i}\mid G_{i}$), and the genotype probability of the $i$th family P($G_{i}$). Each one of these elements are described in Sections \ref{subsec:Multivariate probit model} and \ref{subsec:Genotype probability}.

\subsection{Mixed-effects models for the analysis of family data}
\label{subsec:Multivariate probit model}

To model the correlation of the phenotypes $Y$ and $X$ within families, a common choice is to use random effects. For the binary primary phenotype we propose to use a multivariate probit model with random effects. The advantage of this model is that it involves only the integrals of the multivariate normal cumulative distribution function for which efficient algorithms have been developed. In contrast, for the more commonly used logistic regression model, the integrals have to be approximated for example by using Gauss-Hermite quadrature which might be computationally intensive for large pedigrees. Let $ b^{Y}_{i}=\left(b^{Y}_{i1},\ldots, b^{Y}_{in_{i}} \right)^T$ be a set of family specific random effects designed to handle familial genetic correlation and $ G_{i}=\left(g_{i1},\ldots, g_{in_{i}} \right)^T$ be the vector of genotypes for family $i$. For the probit model, the observed response $Y$ is viewed as a censored observation from an underlying continuous latent variable $Y^{*}$ with:
\begin{equation*}
Y_{ij}=y_{ij} \Leftrightarrow \gamma_{y_{ij}}<Y^{*}_{ij}<\gamma_{y_{ij}+1}, Y_{ij} \in \{0, 1\},
\end{equation*}
where $-\infty$ =$\gamma_0<\gamma_1<\gamma_2=+\infty$ are suitable threshold parameters. For the underlying latent variable $Y^{*}$ we assume the mixed-effects regression model
\begin{equation*}
Y_{i}^{*}=\alpha_{0}+\alpha_{1}G_{i}+\sigma_{G_{Y}}b^{Y}_{i}+\sigma \epsilon^{Y}_{i},
\label{latent variable model}
\end{equation*}
where $\epsilon^Y_{i} \sim N_{n_i}(0,  I_{n_i})$ is independent of $b^{Y}_{i}$. Here $\alpha = (\alpha_0, \alpha_1)$ denotes the regression coefficient vector with $\alpha_0$ the intercept and $\alpha_1$ the parameter representing the effect of the genotype on $Y$. At the family level we assume $b^{Y}_i \sim N_{n_i}(0, \mathbf{R}_i)$, with $\mathbf{R}_i$ the coefficient of relationships matrix with elements $r_{lm} = 2^{-d_{lm}}$ with $d_{lm}$ denoting the genetic distance between subjects $l$ and $m$ in the family. The parameter $\sigma_{G_{Y}}$ represents the residual additive genetic variation not explained by $g_{ij}$. Note that $\sigma_{G_{Y}}$ models the polygenic inheritance in a family.
For identifiability reasons restrictions are required on both the scale and location of $Y^{*}$, namely we set $\sigma^{2}=1$ and $\gamma_1=0$. Thus, in the mixed-effects probit regression the disease risk $\pi_{ij} = P(y_{ij} = 1\mid b^{Y}_{ij}, g_{ij})$ conditional on the random-effects $b^{Y}_{ij}$ and genotypic information $g_{ij}$ is modelled as follows
\begin{equation}
P\left(Y_{ij}=1 \mid g_{ij}, b^{Y}_{ij}\right)=\Phi\left(\alpha_{0}+\alpha_{1}g_{ij}+\sigma_{G_{Y}}b^{Y}_{ij}\right),
\label{Probit model}
\end{equation}
with $\Phi\left(z\right)$ the cumulative distribution function of the standard normal distribution.  The marginal density under the probit model takes the form:
\begin{equation*}
f(y_{ij}\mid g_{ij}; \alpha, \sigma_b) = \int_{b^{Y}_i} \int_{\gamma_{y_{ij}}}^{\gamma_{y_{ij}}+1} f(y^{*}_{ij}\mid g_{ij}, b^{Y}_i; \alpha, \sigma_b) f(b^{Y}_i) dy^{*}_{ij} db^{Y}_i.
\label{Probit marginal distribution}
\end{equation*}
To model the secondary phenotype $X_{i}$ we use a linear mixed model:
\begin{equation}
X_{i}=\beta_{0}+\beta_{1}G_{i}+\sigma_{G_{X}}b^{X}_{i}+\sigma_{\epsilon}\epsilon^{X}_{i},
\label{linear mixed model}
\end{equation}
where $\beta = (\beta_0, \beta_1)$ denotes the regression coefficient vector with $\beta_0$ the intercept and $\beta_1$ the parameter representing the effect of the genotype on $X$, {$b^{X}_{i}$ is the random parameter used to model the genetic correlation structure within each family for the secondary trait, and $\sigma_{\epsilon}$ is the residual standard deviation.

To model jointly $X$ and $Y$ using the model specifications (\ref{Probit model} and \ref{linear mixed model}), we introduce a shared random effect $u_{ij}\sim N(0, 1)$ and propose the following model:
\begin{equation}
\begin{split}
Y_{i}^{*}&=\alpha_{0}+\alpha_{1}G_{i}+\sigma_{G_{Y}}b^{Y}_{i}+\sigma_{u}u_{i}+\epsilon^{Y}_i,\\
X_{i}&=\beta_{0}+\beta_{1}G_{i}+\sigma_{G_{X}}b^{X}_{i}+\delta\sigma_{u}u_{i}+\sigma_{\epsilon}\epsilon^X_{i},
\label{Bivariate model}
\end{split}
\end{equation}
where $u_{i}$ is assumed to be independent of $b^{Y}_{i}, b^{X}_{i}, \epsilon^Y_{i}$, and $\epsilon^X_{i}$. We introduce a coefficient $\delta$ in order to have different phenotypic variances
 for the random effect $u_{i}$. In case of small datasets or small family sizes, it can be better to constrain $\delta$ to be equal to 1 for a simpler model. Let $\Sigma_{X_{i}}$ and $\Sigma_{Y^{*}_{i}}$ denote the corresponding variance-covariance matrices of the marginal distributions of ${X_{i}}$ and ${Y^{*}_{i}}$ and let $\Sigma_{XY^{*}_{i}}$ be their covariance. The joint distribution of the primary and the secondary phenotype is then $\left(Y^{*}_{i},X_{i}\right)\backsim \mathcal{N}_{2n_{i}} \left(\left[ \begin{array}{c} \alpha_{0}+\alpha_{1}G_{i}  \\
\beta_{0}+\beta_{1}G_{i} \end{array}\right], \left[ \begin{array}{cc} \Sigma_{Y^{*}_{i}} & \Sigma_{XY^{*}_{i}} \\
 \Sigma_{XY^{*}_{i}}  & \Sigma_{X_{i}} \end{array}\right]\right)$. In the special case for $n_{i}=2$, the variance-covariance matrix becomes:
\begin{equation}
\Sigma_{i}=
\begin{pmatrix}
\sigma_{G_{Y}}^{2}+\sigma_{u}^{2}+1 & \sigma_{G_{Y}}^{2}2^{-d\left(1,2\right)} & \sigma_{G_{X}}\sigma_{G_{Y}}+\delta\sigma_{u}^{2} & \sigma_{G_{X}}\sigma_{G_{Y}}2^{-d\left(1,2\right)} \\
\sigma_{G_{Y}}^{2}2^{-d\left(1,2\right)} & \sigma_{G_{Y}}+\sigma_{u}^{2}+1 & \sigma_{G_{X}}\sigma_{G_{Y}}2^{-d\left(1,2\right)} & \sigma_{G_{X}}\sigma_{G_{Y}}+\delta\sigma_{u}^{2} \\
\sigma_{G_{X}}\sigma_{G_{Y}}+\delta\sigma_{u}^{2} & \sigma_{G_{X}}\sigma_{G_{Y}}2^{-d\left(1,2\right)} & \sigma_{G_{X}}^{2}+\delta^{2}\sigma_{u}^{2}+\sigma_{\epsilon}^{2} & \sigma_{G_{X}}^{2}2^{-d\left(1,2\right)} \\
\sigma_{G_{X}}\sigma_{G_{Y}}2^{-d\left(1,2\right)} & \sigma_{G_{X}}\sigma_{G_{Y}}+\delta\sigma_{u}^{2} & \sigma_{G_{X}}^{2}2^{-d\left(1,2\right)} & \sigma_{G_{X}}^{2}+\delta^{2}\sigma_{u}^{2}+\sigma_{\epsilon}^{2} \\
\end{pmatrix}.
\label{variance-covariance matrix}
\end{equation}
Using the properties of the multivariate normal distribution, the joint distribution for the observed primary and secondary phenotypes takes the form:
\begin{equation*}
\begin{split}
P\left(Y_{i}, X_{i}\mid G_{i}\right)&=\int P\left(Y_{i}^{*}, X_{i}\mid G_{i}\right)dy_{i}^{*}\\
&=\int P\left(Y_{i}^{*}\mid  X_{i},G_{i}\right)P\left(X_{i}\mid G_{i}\right)dy_{i}^{*}\\
&=P\left(X_{i}\mid G_{i}\right)\int P\left(Y_{i}^{*}\mid  X_{i},G_{i}\right)dy_{i}^{*}.\end{split}
\end{equation*}
Thus by using the probit regression model for the primary trait we have developed an efficient approach to model the correlation between the primary and secondary trait.

From model (\ref{Bivariate model}) and the variance-covariance matrix (\ref{variance-covariance matrix}), several marginal correlations between and within family members can be deduced:
\begin{equation*}
\begin{split}
cor\left(X_{ij},X_{ij'}\right)&=\frac{\sigma_{G_{X}}^{2}2^{-d\left(j,j'\right)}}{\left(\sigma_{G_{X}}^{2}+\delta^{2}\sigma_{u}^{2}+\sigma_{\epsilon}^{2}\right)}=\rho_{X}\\
cor\left(Y_{ij}^{*},Y_{ij'}^{*}\right)&=\frac{2^{-d\left(j,j'\right)}\sigma_{G_{Y}}^{2}}{\left(\sigma_{G_{Y}}^{2}+\sigma_{u}^{2}+1\right)}=\rho_{Y}\\
cor\left(X_{ij},Y_{ij}^{*}\right)&=\frac{\sigma_{G_{X}}\sigma_{G_{Y}}+\delta\sigma_{u}^{2}}{\sqrt{\left(\sigma_{G_{X}}^{2}+\delta^{2}\sigma_{u}^{2}+\sigma_{\epsilon}^{2}\right)\left(\sigma_{G_{Y}}^{2}+\sigma_{u}^{2}+1\right)}}=\rho_{XY}\\
cor\left(X_{ij},Y_{ij'}^{*}\right)&=\frac{2^{-d\left(j,j'\right)}\sigma_{G_{X}}\sigma_{G_{Y}}}{\sqrt{\left(\sigma_{G_{X}}^{2}+\delta^{2}\sigma_{u}^{2}+\sigma_{\epsilon}^{2}\right)\left(\sigma_{G_{Y}}^{2}+\sigma_{u}^{2}+1\right)}}=\rho'_{XY},
\label{formulas correlation}
\end{split}
\end{equation*}
where $\rho_{XY}$ represents the association between the primary and secondary phenotype.
We can also derive the closed form for the heritability estimates of the secondary phenotype which quantifies the percentage of genetic variation in the total variance:
\begin{equation}
H^{2}=\frac{\sigma_{G_{X}}^{2}}{\left(\sigma_{G_{X}}^{2}+\delta\sigma_{u}^{2}+\sigma_{\epsilon}^{2}\right)}.
\label{heritability}
\end{equation}
Note that when genetic factors are included in the model formula (\ref{heritability}) gives the residual heritability.
\subsection{Genotype probability}
\label{subsec:Genotype probability}

Finally another key component in the formulation of the retrospective likelihood (\ref{Likelihood obtained}) is the computation of the genotype probability for each family $i$. Let $G_{mj}$ and $G_{pj}$ denote the genotypes of the mother and father of an individual $j$ if this individual is a nonfounder member of family $i$. Under the assumption of random mating and mendelian inheritance, the genotype probabilities can be written as presented by Duncan Thoms\cite{Thomas2004}:
\begin{equation*}
P\left(G_{i}\right)=\prod_{j=1}^{J} \begin{cases} P\left(g_{ij}\mid G_{mj},G_{pj}\right) & \text{if $j$ is a nonfounder} \\
P\left(g_{ij}\right) & \text{if $j$ is a founder}
\end{cases}.
\label{Genotype probability computation}
\end{equation*}
 The probabilities $P(g_{ij}\mid G_{pj}, G_{mj})$ are the transmission probabilities which can be modelled using mendelian inheritance. Finally $P\left(G_{pi}\right)$, $P\left(G_{mi}\right)$, and $P\left(g_{ij}\right)$ can be modelled by assuming Hardy-Weinberg proportions $\left(1-q\right)^{2}$, $2q\left(1-q\right)$, $q^{2}$ which depend on $q$, the minor allele frequency. Here we propose to use external information for $q$ or to estimate $q$ from the control sample before maximizing the likelihood. Note that when genotypes of the parents are missing the probability can be obtained by summing over the possible parental genotypes. In case of more complex pedigree a recursive algorithm known as peeling (Elston and Stewart, 1971)
can be used \cite{Elston1971}. For the LLS where families are sibships the probability is as follows:
\begin{equation}
L\left(\theta; Y,X\right)=\prod_i \frac{\left\{P\left(X_{i}\mid G_{i}\right)\int P\left(Y_{i}^{*}\mid  X_{i},G_{i}\right)dy_{i}^{*}\right\} \sum_{G_{p}}\sum_{G_{m}}\prod_{j}P\left(G_{ij}\mid G_{m},G_{p}\right)P\left(G_{p}\right)P\left(G_{m}\right)}{\sum_{g}\sum_{G_{p}}\sum_{G_{m}}\int P\left(Y_{i}^{*}\mid g\right)P\left(g\mid G_{m},G_{p}\right)P\left(G_{p}\right)P\left(G_{m}\right)},
\label{Log-likelihood}
\end{equation}
where $\theta = (\alpha_{0}, \alpha_{1}, \sigma_{G_{Y}}, \beta_{0}, \beta_{1}, \sigma_{G_{X}}, \sigma_{\epsilon}, \delta, \sigma_{u})$ is the model parameters vector.
\subsection{Estimation and statistical testing}

To estimate the parameters of the joint model we maximize the logarithm of the likelihood described in (\ref{Likelihood obtained}). This involves a combination of numerical optimization and integration. For the evaluation of the integral in the multivariate normal distribution, we use the deterministic algorithm Miwa described in \cite{Miwa2003}. For the optimization, we use the Broyden-Fletcher-Goldfarb-Shanno (BFGS) algorithm implemented in the function \texttt{optim(.)} in R. The BFGS algorithm is a quasi-Newton method, which means that the Hessian matrix does not need to be evaluated directly but is approximated by using specified gradient evaluations. To test for the presence of an effect of the SNPs on the secondary phenotype we use the likelihood ratio test. Note that when the interest of a researcher is solely testing for genetic association a score statistic can might be an alternative for the likelihood ratio statistic.

\subsection{Continuous polygenic score}

Our approach can also be applied in the case of modelling the association between continuous covariates and secondary phenotypes. For example polygenic scores have been used to summarise genetic effects among an ensemble of SNPs that have been identified in large GWASes \cite{Purcell2009, Bush2010, Simonson2011}. Polygenic scores are typically linear combinations of SNPs: $G=\sum_k \delta_{k}SNP_{k}$, where $\delta_{k} = 1$ or $\delta_{k}$ is obtained from previous GWASes. For genetic scores, we need to integrate over the distribution of the polygenic score instead of summing over the genotypes in the denominator of (\ref{Likelihood obtained}). For the distribution of the polygenic score we use a multivariate normal distribution $G_{i} \backsim \mathcal{N}_{n_{i}} \left(\mu_{g}, \sigma_{g}R_{i}\right)$, with $\mu_{g}$ the mean value of the genetic score, $\sigma_{g}$ the standard deviation of the genetic score and $R_{i}$ the relationship matrix of family $i$. The likelihood contribution for family $i$ is given by:
\begin{equation*}
\frac{P\left(Y_{i},X_{i}\mid G_{i}\right)P\left(G_{i}\right)}{P(Y_{i})}=\frac{P\left(Y_{i},X_{i}\mid G_{i}\right)P\left(G_{i}\right)}{\int_{y_{i}^{*} }P(y_{i}^{*})dy_{i}^{*}}=\frac{P\left(Y_{i},X_{i}\mid G_{i}\right)P\left(G_{i}\right)}{\int_{y^{*}}\int_{g_{i} }P(y_{i}^{*}\mid g_{i})P(g)dy_{i}^{*}dg_{i}}.
\end{equation*}
Computation of the integral $\int_{y^{*} }\int_{g}P(y^{*}\mid g)P(g)dy_{*}dg$ can be quite intensive and challenging. In order to gain efficiency we write the marginal model of $Y^{*}$ (\ref{Bivariate model}) as $Y_{i}^{*}=\alpha_{0}+b_{i}^{Y*}+u_{i}+\epsilon^{Y}_{i}$, with $b_{i}^{Y*}=\sigma_{G_{Y}}b^{Y}_{i}+\alpha_{1}G_{i}$.
Now $Y_{i}^{*}$ follows the following multivariate normal distribution: $Y_{i}^{*} \backsim \mathcal{N}_{n_{i}} \left(\alpha_{0}+\alpha_{1}\mu_{g}, \Sigma_{Y^{*}_{i}}+\alpha_{1}^{2}\sigma_{g}^{2}R_{i}\right)$. Note that when a polygenic risk score is included in the model for the secondary phenotype, the parameter $ \sigma_{G_{Y}}$ represents the residual polygenic inheritance.

\subsection{Inclusion of covariates in the model}
Often, researchers want to adjust for covariates such as age, sex, treatment etc in the model. Let $Z$ be such a covariate. To estimate the effect $Z$ on the secondary phenotype we propose to maximize the joint likelihood of $X$ and $G$ conditionally on the primary phenotype $Y$ and $Z$. Thereby we avoid modeling of the distribution of $Z$ within the families. Indeed, under the assumption of independence between genotype and $Z$ we obtain:

\begin{equation}
P\left(X_{i},G_{i}\mid  Y_{i},Z_{i}\right)=\frac{P\left(X_{i},Y_{i},Z_{i} ,G_{i}\right)}{P\left(Y_{i},Z_{i}\right)}=\frac{P\left(X_{i},Y_{i}\mid G_{i},Z_{i}\right)P\left(G_{i}\right)P\left(Z_{i}\right)}{P\left(Y_{i}|Z_{i}\right)P\left(Z_{i}\right)}=\frac{P\left(X_{i},Y_{i}\mid G_{i},Z_{i}\right)P\left(G_{i}\right)}{P\left(Y_{i}|Z_{i}\right)}.
\label{Covariates}
\end{equation}

\section{Simulation Study}
A simulation study has been set up to evaluate the performance of our proposed method for the estimation of the association between a genetic factor and the secondary phenotype and the estimation of the heritability of the secondary phenotype. We compare our method with the naive approach which is typically followed in practice, namely analysis of the secondary trait without correcting for the sampling mechanism. In particular, in this case, we fit the standard linear mixed-effects model for the secondary phenotype and explicitly model the familial relationships as described in (\ref{linear mixed model}).  The two methods are compared in terms of bias, Root Mean Square Error (RMSE) and  95\% coverage probabilities. We consider SNPs (discrete variables) and polygenic scores (continuous variables). Several settings are considered for the disease prevalence, the strength of the association between the genetic factor and the primary phenotype, the strength of the ascertainment mechanism and the number of families. We use families of size 5.  With respect to the familial relationships, we consider only sibships such that our simulation resembles the LLS design. For the prevalence of the primary phenotype we consider two settings namely  a disease prevalence of 1\% which corresponds to $\alpha_{0} \approx -2.32 $ and of 5\% which corresponds to $\alpha_{0} \approx -1.64 $. In addition the variance parameters have been chosen such that they correspond to a heritability of 50\%. Specifically  we use $\sigma_{G_{X}}$=2, $\sigma_{G_{Y}}=\sqrt{3}$, $\sigma_{u_{X}}=\sigma_{u_{Y}}=\sqrt{2}$ and $\sigma_{\epsilon} =\sqrt{2}$. This corresponds to a correlation of 0.78 between the primary and the secondary phenotypes. To speed up computations, we assume that $\sigma_{u_{X}}=\sigma_{u_{Y}}$ when fitting the models to the simulated datasets. For each scenario, 500 datasets are simulated using model (\ref{Bivariate model}).

\subsection{Simulation results for a SNP}

The genotypes of the SNPs are simulated assuming a minor allele frequency of 0.3 in the population. For the secondary phenotype model the following fixed effects values are used: $\beta_{0}= 3.5 $ and $\beta_{1}= 0.2 $, whereas for the primary phenotype model the effect sizes are $\alpha_{1} = $ 0.1 or 0.5. Finally, for each of the four scenarios (rare or common disease, and weak and strong SNP effect on the primary phenotype) we consider two ascertainment mechanisms, namely the sampled families of size five have at least one affected and at least two affected members.

Figure \ref{Fig1} presents the estimates and 95\% confidence intervals for the scenario of 400 families. Figure \ref{Fig1} shows that ignoring the sampling mechanism (naive method) leads to biased estimates of the SNP effect and the size of this bias increases with the strength of the ascertainment mechanism and the association between the SNP and the primary phenotype. Overall we observe that our method gives unbiased estimates of the SNP effect on the secondary phenotype. The coverage probabilities reach the nominal level (see section A of supplementary material). Regarding the prevalence of the primary phenotype, we observe that for the naive method bias increases with lower prevalence, while our method remains robust to the lower amount of information due to the rare primary phenotype. In general, our method leads to smaller RMSE than the naive approach and better coverage probabilities.

In Table \ref{Heritability} we present the heritability estimates of the secondary phenotype  for a common disease, under the various ascertainment mechanisms and the two values of $\alpha_{1}$. It is obvious that the heritability estimates are influenced by the ascertainment mechanisms when using the naive approach. Indeed the naive method tends to underestimate the heritability for each mechanism and this underestimation is getting larger as the ascertainment mechanisms become more stringent. The heritability estimates are 25-27\% for families with at least one affected sibling and drop to 13-14\% for families with at least 2 affected siblings. On the contrary, our method is robust to the stringency of the ascertainment mechanism.

Next, we investigate the performance of our approach in terms of type I error in each of the four considered scenarios. We simulate 10,000 replicates for each scenario. In Table \ref{Type I error} the type I error rates are given for the rare disease scenario (i.e. prevalence 1\%). We observe that while our approach preserves the type I error rate at a nominal level, the naive approach has, systematically, an inflated type I error rate. The type I error rate for the naive method increases with stronger ascertainment and larger SNP effect on the primary phenotype.

Finally, we study the robustness of our approach to one violation of the model assumptions, namely we simulated under a logit link for the primary phenotype and used the probit link for modelling.  Results for the SNP effect and the heritability are presented in Table \ref{Robustness}. These results show that even though our approach gives biased estimates for the primary phenotype model, the parameters estimates for the secondary phenotype model are not affected. All the results are presented in Section A of the Supplementary Material.

\subsection{Simulation results for a polygenic score}
To study the performance of our method for polygenic score, we simulated centered and standardized scores. The parameters of the secondary phenotype model were chosen as for the SNP simulations: $\beta_{0}= 3.5 $ and $\beta_{1}= 0.2 $, whereas for the primary phenotype model effect sizes of $\alpha_{1} = $ 0.1 or 0.5 were used. Figure \ref{Fig3} presents the estimates and confidence intervals for datasets with 400 families. Our approach provides unbiased estimates of the effect of the polygenic score on the secondary phenotype. In contrast, the naive approach provides biased estimates and the bias increases when the ascertainment process is more stringent or when $\alpha_1$ is larger.

In Table \ref{Heritability} the simulations results with regard to the heritability estimates are given. The results of the residual heritability estimates after adjustment for polygenic scores agree with the results obtained when a SNP is included in the model. The naive approach did not perform well: estimates between 25-26\% and 14-15\% for an ascertainment process of at least one affected sibling and at least two affected siblings respectively instead of 50\%.

\section{Application: Analysis of the Leiden Longevity Study}

In this Section, we will exemplify our proposed method in the analysis of the LLS which has been briefly introduced in Section 1. The LLS is a family-based study which has been set up to identify mechanisms that contribute to healthy ageing and longevity. The inclusion criteria of the study are families with at least two nonagenarian siblings, i.e. the selection takes place at Generation II (Figure \ref{LLS pedigree}). Several secondary phenotypes and GWAS data have been measured for the offspring of these siblings (Generation III in Figure \ref{LLS pedigree}) and their partners. Since the offspring have at least one nonagenarian parent, they are also likely to become long-lived. Therefore, the set of offspring and their partners corresponds to a case-control design with related subjects where the offspring in Generation III are considered as cases and their partners as controls. Overall 421 families with 1671 offspring (cases) and 744 partners (controls) have been included in the study. Because the families are relatively small we use the model which assumes an equal variance for the shared effect for the two traits.

Here we model the association between genetic factors and the secondary phenotypes triglyceride and glucose levels. For both traits, there is evidence of an association with human longevity and both traits are normally distributed. For the sake of comparison in addition to our proposed method, we will present results using the naive approach i.e. standard linear mixed model. Analyses using the linear mixed model which conditions also on the case-control status will not be presented because the parameters do not have a comparable interpretation between the two approaches.

\subsection{Triglyceride levels analysis}

Triglyceride levels have been found to be associated with the primary trait longevity ($p$-value = 0.0005 for women and $p$-value = 0.04 for men) and the size of association is sex dependent. Therefore a sex-stratified analysis has been considered further. For the purposes of our illustration, we restricted our analysis to seven genes on chromosome 11 which are known to be associated with Triglyceride levels. These genes are \textit{APOA1, APOA4, APOA5, APOC3, ZNF259, BUD13} and \textit{DSCAML1}. For these genes, we have genotypes of 41 SNPs which have no missing values in our datasets. Triglyceride levels were standardized and we included age as a covariate in the analysis.

We ran the analysis with the constrained approach, i.e. $\delta = 1$. We observe that none of the SNPs analysed is significantly associated with Triglyceride levels neither in men nor in women, hence for most SNPs the estimates of the effect sizes agree between the two approaches. The SNPs showing the largest differences are, in men, SNP 22: $\beta^{RA}_{1}$ = 0.047 for our Retrospective Approach (RA) and $\beta^{NA}_{1}$ = 0.052 for the Naive Approach (NA) and SNP 26: $\beta^{RA}_{1}$ = 0.088 and $\beta^{NA}_{1}$ = 0.092. For women more SNPs give different estimates between the two approaches, i.e. SNP 1 ($\beta^{RA}_{1}$ = 0.024, $\beta^{NA}_{1}$ = 0.020), SNP 2 ($\beta^{RA}_{1}$ = 7.2e-06 $\beta^{NA}_{1}$=0.006), SNP 13 ($\beta^{RA}_{1}$ = -0.013, $\beta^{NA}_{1}$ = -0.009) and SNP 19 ($\beta^{RA}_{1}$ = 0.011, $\beta^{NA}_{1}$ = 0.007) showed the biggest differences. Results for the SNPs are presented in Section B of the Supplementary Material.

We verified whether the assumption of equal variances for the primary and secondary phenotype for the shared effects is justified. We fitted also the model with non constrained $\delta$. We noticed that for some of the SNPs the model parameters are hard to estimate and a switch between the estimates of the variances of the shared and residual random effects in the model for the second phenotype seems  to occur. Overall the estimates of the effect of the SNP on the secondary phenotype are very similar to the model which assumes equal variances. Results of these analyses are presented in Section B of the Supplementary Material.

\subsection{Glucose levels analysis}

We now proceed with the analysis of glucose levels in the offspring and partners of the LLS. Mooijart \cite{Mooijaart2010} studied the association between glucose and a polygenic score. The genetic score was defined as the total number of risk alleles across 15 SNPs which are known to be associated with Type II diabetes. The Generalized Estimating Equation method was applied to take into account the familial relationships. The paper showed that a higher number of Type II diabetes risk alleles is associated with a higher serum concentration of glucose ($p-value$ = 0.016). A statistically significant association was found between glucose level and case-control status (p-value$<$ 0.001). However, the sampling process was not taken into account in the analysis and thus the results might be biased. We applied our method to estimate the heritability of glucose levels and to test for the presence of an association between the glucose levels and the polygenic score. In addition, we applied the naive approach which did not correct for case-control status. We did not stratify according to sex in these analyses.

For this analysis the polygenic score was standardized. Using the Retrospective approach, the association between the genetic score and the glucose level is estimated by $\beta^{RA}_{1} = 0.630$ with a standard error of  $stE = 0.023$ ($p-value = 0.015$). The naive approach also yields a significant association between the genetic score and glucose levels but with a different $\beta$ ($\beta^{NA}_{1} = 0.622$, $stE = 0.026$, $p-value = 0.020$). By using the Naive Approach (NA) we obtained for the glucose levels a genetic variance of $\sigma_{G_{X}}^{2} = 0.302$ and a total variance of $\sigma_{T}^{2} = 1.322$, which corresponds to a residual heritability of $h^{2}_{NA} = 0.228$. Our Retrospective approach (RA) yields a genetic variance of $\sigma_{G_{X}}^{2} = 0.384$ and a total variance of $\sigma_{T}^{2} = 1.457$ which corresponds to a residual heritability of $h^{2}_{RA} = 0.263$.

\section{Discussion}

In this paper, we developed a new method for the proper analysis of secondary traits for multiple-cases family designs. A key component in our proposed method is the joint modelling of the primary and secondary phenotypes. We developed a multivariate probit model which can also capture the within families dependencies. A retrospective likelihood approach has been followed to correct for the ascertainment process. Thereby unbiased estimates of the association between genetic factors and secondary traits can be obtained. Simulation results showed that our approach preserves the type I error at nominal level and provides accurate estimates irrespective of the disease prevalence, the strength of the association between the genetic variants and the primary phenotype, and the ascertainment mechanism.  Another important empirical finding is that the heritability estimates for the secondary traits can be severely underestimated unless the sampling mechanism is taken into account. With respect to the analysis of the motivating case study, for the SNPs the differences between the effect sizes obtained by our proposed method and the naive approach were small. However typically many SNPs are considered and small differences might, therefore, be important. With regard to the effect size of the genetic score and the residual heritability of glucose, the difference between our approach and the naive approach was larger.

Heritability is one of the properties that a trait needs to possess to be declared an endophenotype for a specific disease. The other criteria are: the trait is associated with the disease status in the population, the trait is primarily state-independent, and the trait and the disease status co-segregate within a family \cite{Gottesman2003}. The Leiden Family Lab (https://www.leidenfamilylab.nl) aims to identify endophenotypes for social anxiety disorder. The study comprises families with at least two cases with social anxiety. The methods presented in this paper will be used for the analyses of this study to identify endophenotypes and are relevant for other family studies, as well.

In this paper, we proposed to include additional covariates in the model by using the likelihood conditional on these covariates. Alternatively the joint likelihood of the secondary phenotype, genotype, and covariate conditionally on the primary phenotype can be used. This alternative approach might be more efficient \cite{Balliu2015}. However this likelihood requires distributional assumptions for the covariates within families which can be complex for related individuals. Moreover maximization of the likelihood might become time consuming.  Ghosh \emph{et al} \cite{Ghosh2013} propose a pseudo-likelihood and a profile approach to include covariates in a secondary phenotype analysis for case-control data. This work needs to be extended to family data. A Monte Carlo approach might be considered to compute the integrals (Tsonaka \emph{et al} \cite{Tsonaka2015}). This is a topic for future research.

Typically there are missing genotypes. In unrelated individuals, genotypes can be imputed based on the haplotype structure obtained from a reference panel. For family data, the imputation should also take into account the genotypes of other family members. Software exists which can perform such analysis, for example the Genotype Imputation Given Inheritance (GIGI) program \cite{Cheung2013}. However for the computation of the denominator in equation (\ref{Likelihood obtained}) these imputed genotype probabilities have to be taken into account. How to incorporate them is a topic for future research.

Due to the computational intensity of the proposed method, it is not yet possible to run full GWAs analyses of secondary phenotypes. However, our method can be used on a set of pre-selected variants e.g. after an initial screening with the naive approach to the primary and secondary phenotypes or when investigating pleiotropic effects. For fastest computation of the multivariate integrals in the numerator and the denominator, a faster algorithm can be used than the one used in this paper. The randomized Quasi-Monte-Carlo procedure, developed by Genz \cite{Genz1992}, is less accurate but faster especially for large pedigrees. Development of less computational intensive methods is one of the topics for future research.

With regard to pleiotropic effects, a criticism of probit random-effects models is that in the presence of high dimensional random effects we cannot move from the subject-specific interpretation for the fixed effects parameters to the population-level interpretation as in the random-intercepts case. Note that when the binary outcome of interest and when we have relatively small family studies, the intercept and variance terms might be hard to estimate due to lack of information and poor coverage probabilities are observed. Tsonaka \emph{et al} \cite{Tsonaka2012} showed that by using information on the prevalence of the disease efficiency can be gained. Their methods need to be adapted to our setting of the analysis of two phenotypes. When the parameters of the primary phenotype model are not of interest and this model is only used to correct for the ascertainment mechanism which is driven by the primary phenotype, we showed that secondary phenotype analyses with our method are robust to using the probit instead of the logit link function.

Several other extensions and future directions can be followed. First, in the LLS and the Leiden Family Lab Study several omics and fMRI data will become available in the near future, respectively, and joint modelling of several glycans or voxels is of interest. Extending our approach, in this case, is algebraically straightforward, but it might be hard to implement in practice due to its computational intensity especially with a high number of secondary phenotypes. Use of composite likelihood approaches might be a solution and is one of our future research topics.

Finally, an attractive alternative approach to properly analyse secondary traits is to apply inverse probability weighting. However, it is crucial to correctly specify the weights. Currently, we do not have sufficient information to be able to estimate these weights for our studies. However with the availability of electronic health records, such as information from general practitioners, for research, we might have access to the information needed to estimate the weights and inverse probability weighting approaches can be developed.

\section*{Acknowledgments}
This work was supported by FP7-Health-F5-2012, under grant agreement n$^{o}$305280 (MIMOmics).

\bibliographystyle{wileyj}
\bibliography{refs}

\clearpage

\begin{table}
\caption{Type I errors rates for testing for association between a genetic marker and a secondary phenotype for four scenarios. Families with at least one and with at least two cases are considered. Two values for the association between the SNP and the primary phenotype namely $\alpha_{1}$ = 0.1 and $\alpha_{1}$ = 0.5 are used. Datasets consist of 400 families of size 5. Results are based on 10000 replicates.   \label{Type I error}}
\centering
\begin{tabular}{ l r c c}
\hline
 & nominal level ($\alpha$) & Retrospective likelihood &  Naive method \\
 \hline
At least 2 cases  &  & &\\
$\alpha_{1}$ =0.1 & & &\\
& 0.05  & 0.0509  & 0.0580 \\
 & 0.01 & 0.0118  & 0.0152 \\
 & 0.001 & 0.0017  & 0.0025  \\
$\alpha_{1}$ =0.5 & & &\\
& 0.05 & 0.0505  & 0.0878  \\
& 0.01 & 0.0113  & 0.0222  \\
& 0.001 & 0.0013 & 0.0043  \\[4pt]
At least 1 case  &  & &\\
$\alpha_{1}$ =0.1  & & &\\
& 0.05 & 0.0524& 0.0514   \\
 &0.01 & 0.0102 & 0.0098  \\
 & 0.001 & 0.0018 & 0.0014 \\
$\alpha_{1}$ =0.5  & & &\\
& 0.05 & 0.0522 & 0.0558 \\
& 0.01 & 0.0098 & 0.0097 \\
& 0.001 & 0.0009 & 0.0016 \\
\hline
\end{tabular}
\end{table}

\begin{table}
\caption{Heritability results of the simulation studies for a SNP and a polygenic score: Estimates with standard deviations and RMSE (in brackets) for the heritability of the secondary phenotype for a common disease (prevalence $\approx$ 5\%), when families with at least one and at least two cases are sampled and for two values of $\alpha_{1}$, i.e. SNP or polygenic score effect on primary phenotype. Datasets consist of 400 families of size 5. Results are based on 500 replicates. \label{Heritability}}
\centering
\begin{tabular}{ l c c c c c }
\hline
& & \multicolumn{2}{c}{SNP model} & \multicolumn{2}{c}{Polygenic score model} \\
Ascertainment & $\alpha_{1}$ & Retrospective & Naive & Retrospective & Naive \\
\hline
1. 2 cases & & & \\[2pt]
& 0.10 & 0.48(0.07)(0.22) & 0.13(0.07)(0.37) & 0.50(0.03)(0.13) & 0.14(0.03)(0.36)\\
& 0.50 & 0.48(0.07)(0.22) & 0.14(0.07)(0.36) & 0.52(0.03)0.12) & 0.15(0.03)(0.34)\\

2. 1 case & & & \\[2pt]
& 0.10 & 0.50(0.08)(0.17) & 0.25(0.08)(0.25) & 0.48(0.04)(0.12) & 0.25(0.03)(0.24)\\
& 0.50 & 0.50(0.08)(0.17) & 0.27(0.08)(0.24) & 0.50(0.04)(0.10) & 0.26(0.04)(0.23)\\
\hline
\end{tabular}
\end{table}

\begin{table}
\caption{Robustness: Estimates of the effect size of the SNP on the secondary phenotype ($\beta_{1}$) and heritability of the secondary phenotype are given for a common disease (prevalence $\approx$ 5\%), for the two ascertainment mechanisms and two values of $\alpha_{1}$. Into brackets are standard deviations, RMSE and coverage probability (for the effect size only). Datasets consist of 400 families of size 5. Results are based on 500 replicates. \label{Robustness}}
\centering
\begin{tabular}{ l  c  c c }
\hline
Ascertainment & $\alpha_{1}$ & $\beta_{1}$ & heritability\\
\hline
0.True value & & 0.200 & 0.500\\[2pt]

1.At least 2 cases & & & \\[2pt]
& 0.100 & 0.199(0.104)(0.104)(0.948) & 0.509(0.017)(0.110)\\
& 0.500 & 0.197(0.106)(0.110)(0.945) & 0.516(0.014)(0.108)\\

2.At least 1 case & & & \\[2pt]
& 0.100 & 0.200(0.104)(0.107)(0.961) & 0.510(0.012)(0.096)\\
& 0.500 & 0.199(0.107)(0.111)(0.960) & 0.513(0.010)(0.087)\\
\hline
\end{tabular}
\end{table}

\begin{figure}[h]
\centering\includegraphics[scale=0.6]{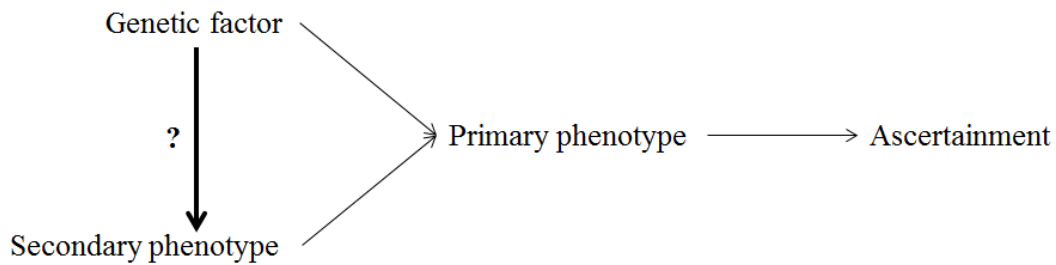}
\caption{Directed acyclic graph representing the case where bias is expected when estimating the association between the genetic marker and the secondary phenotype. Arrows represent existing association between each node of the graph. A secondary phenotype analysis investigates whether there is an association between the genetic factor and the secondary phenotype}
\label{DAG}
\end{figure}

\begin{figure}[h]
\centering\includegraphics[scale=0.3]{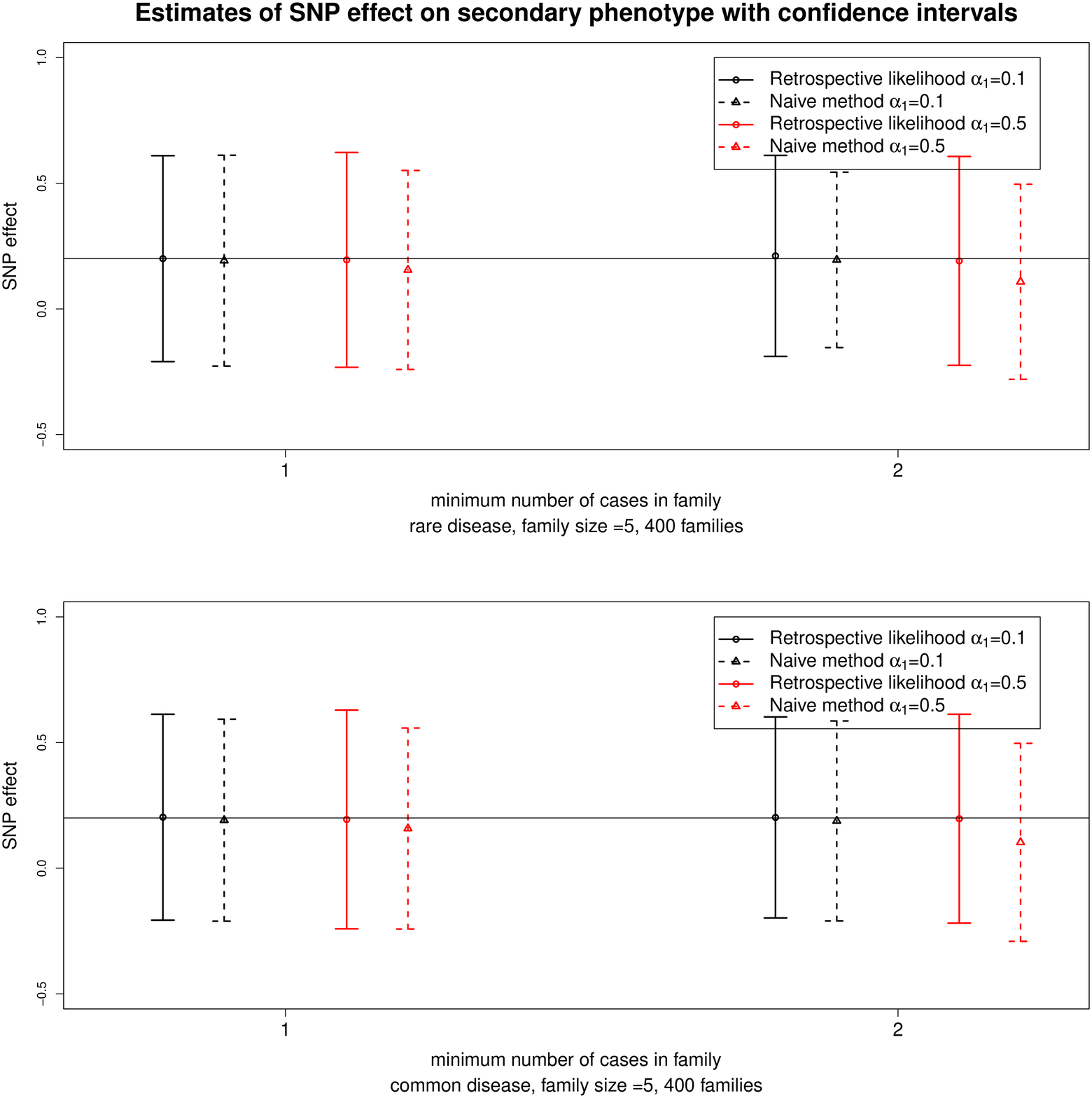}
\caption{Estimates and 95\% confidence intervals for the SNP effect on the secondary phenotype for the retrospective likelihood approach and the naive method. Results are obtained from 500 simulated datasets of 400 families for 2 ascertainment schedules. The top and bottom panel correspond to a rare or common primary phenotype with a prevalence around 1\% and 5\% respectively. In black and red are represented results for small ($\alpha_1$=0.1) and large ($\alpha_1$=0.5) effect sizes of the SNP on the primary phenotype, respectively. The horizontal line corresponds to the true SNP effect on the secondary phenotype.}
\label{Fig1}
\end{figure}

\begin{figure}[h]
\centering\includegraphics[scale=0.3]{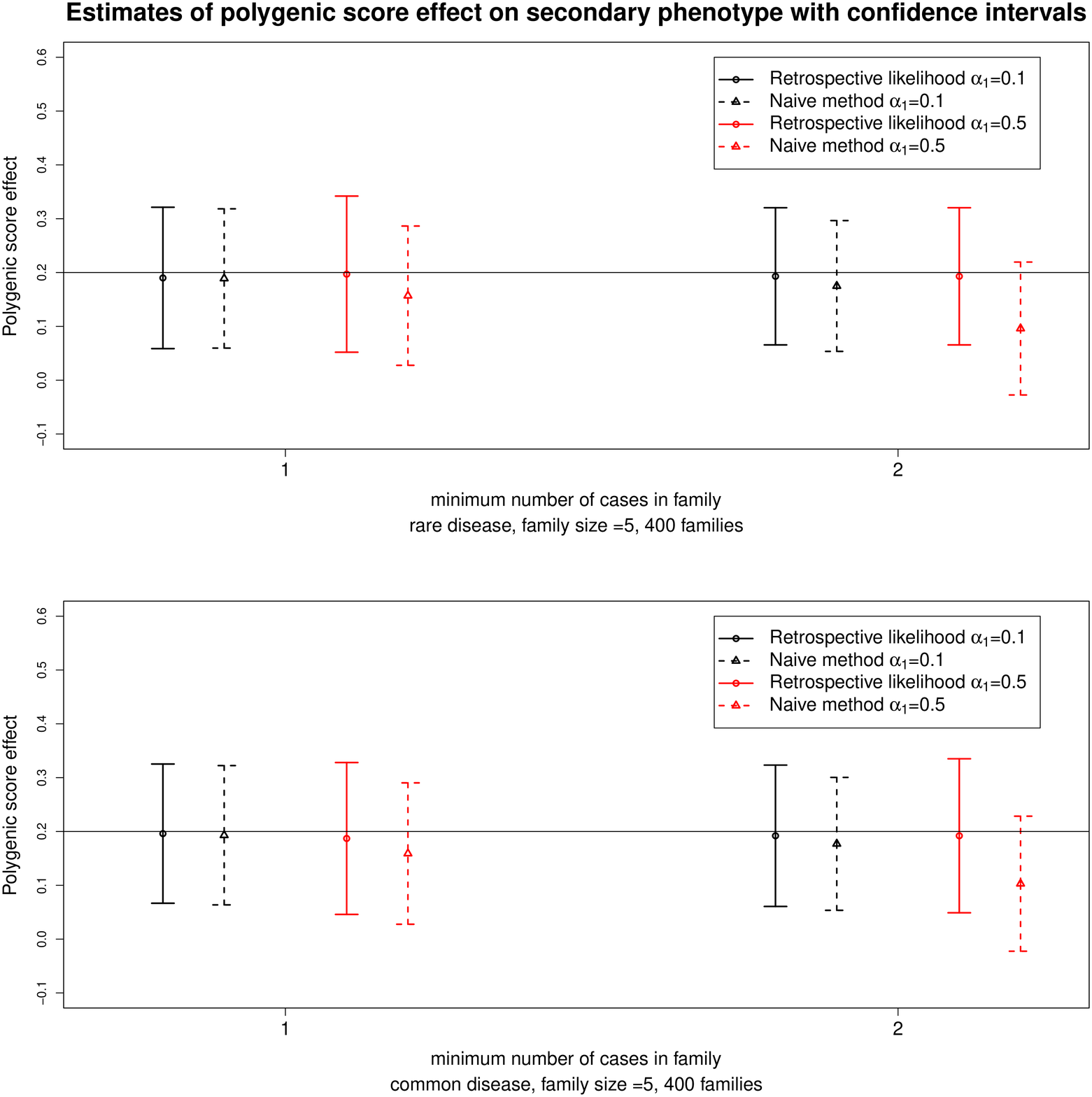}
\caption{Estimates and 95\% confidence intervals for the polygenic score effect on the secondary phenotype for the retrospective likelihood approach and the naive method. Results are obtained from 500 simulated datasets of 400 families for 2 ascertainment schedules. The top and bottom panel correspond to a rare or common primary phenotype with a prevalence around 1\% and 5\% respectively. In black and red are represented results for small ($\alpha_1$=0.1) and large ($\alpha_1$=0.5) effect sizes of the polygenic score on the primary phenotype, respectively. The horizontal line corresponds to the true polygenic score effect on the secondary phenotype.}
\label{Fig3}
\end{figure}

\begin{figure}[h]
\centering\includegraphics[scale=0.4]{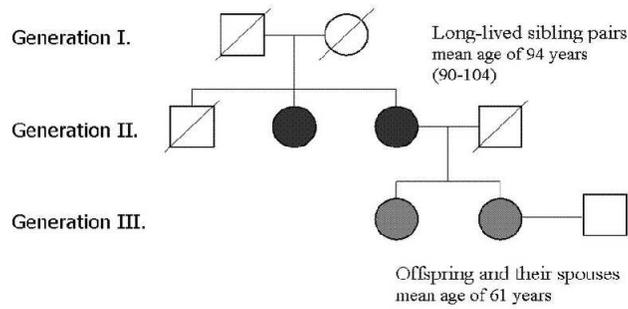}
\caption{Example of a family pedigree from the LLS. Squares and circles represent men and women respectively, crossed symbols represent deceased individuals. In black are the long-lived individuals on whom the ascertainment is based, in grey are the cases of the study (offsprings of long-lived siblings) and in white are the controls.}
\label{LLS pedigree}
\end{figure}

\clearpage
\newpage
\beginsupplement

\section{Supplementary Material for: Secondary phenotype analysis in ascertained family designs: Application to the Leiden Longevity Study}

In this supplementary materials, we are presenting all the simulations results obtained to compare our restrospective likelihood approach to the naive mixed model approach in Section A. In section B are presented the association results between 41 selected SNPs and triglyceride levels obtained on the Leiden Longevity Study.

\subsection{Simulation results}
\subsubsection{Description of the simulation study}
A simulation study has been set up to evaluate the performance of our proposed method in various settings for the correlation between the two outcomes, the disease prevalence, the strength of the association between the SNP and the primary phenotype and strength of the ascertainment mechanism. In addition, we contrasted our method with the naive approach which is typically followed in practice, namely analysis of the secondary trait without correcting for the sampling mechanism. In particular in this case we applied the standard linear mixed effects model for the secondary phenotype and explicitly modelled the familial relationship. The two methods have been compared in terms of Root Mean Square Error (RMSE), and 95\% coverage probabilities. We simulate multiple cases family data and secondary phenotypes for sibships using the mixed-effects logistic regression:
\begin{equation}
\begin{split}
Y_{i}^{*}&=\alpha_{0}+\alpha_{1}G_{i}+\sigma_{G_{Y}}b^{Y}_{i}+\sigma_{u}u_{i}+\epsilon_{Yi}\\
X_{i}&=\beta_{0}+\beta_{1}G_{i}+\sigma_{G_{X}}b^{X}_{i}+\sigma_{u}u_{i}+\sigma_{\epsilon}\epsilon_{Xi}
\end{split}
\end{equation}
With respect to the familial relationships we have simulated families with only siblings such that our simulation resembles more the LLS design. For the prevalence of the primary phenotype we considered two cases: the disease prevalence equals 1\% which corresponds to $\alpha_{0} \approx -2.32 $ and 5\% which corresponds to $\alpha_{0} \approx -1.64 $. The SNP effect on the primary phenotype measured by $\alpha_{1}$, was taken equal to $0.1$, $0.5$ or $1$. In addition the remaining variance parameters have been chosen such that they correspond to 50\% heritability, i.e. ,$\sigma_{G_{X}}$=2, $\sigma_{G_{Y}}=\sqrt{3}$, $\sigma_{u}=\sqrt{2}$ and $\sigma_{\epsilon} =\sqrt{2}$.  which corresponds to 0.78 correlation between the primary and the secondary phenotypes. Finally for the secondary phenotype we choose as fixed effects values :$\beta_{0}= 3.5 $ and $\beta_{1}= 0.2$ or $2$. Finally for each of the 4 scenarios (rare or low disease and low and higher SNP effect on the primary phenotype) we considered 4 ascertainment mechanisms, i.e. we assumed that families have been sampled provided that at least 1 or 2 out of the 5 members are affected. For each dataset 400 families were simulated.

\subsubsection{Simulation study results for a SNP as genetic marker}

\begin{table}[h]
\begin{center}
\caption{Simulations results obtained on 500 datasets with 400 families of size 5 with at least 2 cases of a rare disease (prevalence around 1\%) and $\alpha_{1}$=0.1. Into brackets are respectively standard deviations, root mean square errors and the coverage probabilities.}
\scalebox{0.7}{
\begin{tabular}{r c c c}
\hline
 & Real Value & Retrospective likelihood & Naive method \\
 && Est (SD) (RMSE) (Cov Pr) & Est (SD) (RMSE) (Cov Pr)\\
\hline
 $\beta_0$ & 3.500 & 3.612(1.444)(1.533)(0.859) & 5.192(0.178)(1.702)(0.000) \\
 $\beta_1$ & 0.200 & 0.211(0.204)(0.212)(0.936) & 0.195(0.215)(218)(0.934) \\
 $\sigma_{GX}$ & 2.000 & 1.841(0.479)(0.795)(0.622) & 0.913(0.583)(1.131)(0.128)\\
 $\sigma_{\epsilon}$ & 1.414 & 1.552(0.145)(0.271)(0.662) & 2.471(0.049)(1.063)(0.000) \\
 $\alpha_0$ & -2.326 & -2.279(1.650)(2.112)(0.773) &- \\
 $\alpha_1$ & 0.100 & 0.087(0.176)(0.196)(0.906) & - \\
 $\sigma_{GY}$ & 1.732 & 1.317(0.677)(0.965)(0.844) & - \\
 $\sigma_{u}$ & 1.414 & 0.995(0.687)(0.778)(0.730) & - \\
   \hline
\end{tabular}}
\end{center}
\end{table}

\begin{table}[h]
\begin{center}
\caption{Simulations results obtained on 500 datasets with 400 families of size 5 with at least 2 cases of a rare disease (prevalence around 1\%) and $\alpha_{1}$=0.5. Into brackets are respectively standard deviations, root mean square errors and the coverage probabilities.}
\scalebox{0.7}{
\begin{tabular}{r c c c}
\hline
 & Real Value & Retrospective likelihood & Naive method \\
 && Est (SD) (RMSE) (Cov Pr) & Est (SD) (RMSE) (Cov Pr)\\
\hline
 $\beta_0$ & 3.500 & 3.678(1.140)(1.172)(0.932) & 5.099(0.185)(1.610)(0.000) \\
 $\beta_1$ & 0.200 & 0.191(0.212)(0.228)(0.906) & 0.108(0.198)(0.234)(0.893) \\
 $\sigma_{GX}$ & 2.000 & 1.899(0.561)(0.693)(0.728) & 0.974(0.458)(1.068)(0.106) \\
 $\sigma_{\epsilon}$ & 1.414 & 1.529(0.123)(0.231)(0.707) & 2.444(0.051)(1.038)(0.000) \\
 $\alpha_0$ & -2.326 & -1.989(1.449)(1.814)(0.793) & - \\
 $\alpha_1$ & 0.500 & 0.416(0.173)(0.262)(0.905) & - \\
 $\sigma_{GY}$ & 1.732 & 1.461(0.673)(1.044)(0.876) & - \\
 $\sigma_{u}$ & 1.414 & 1.072(0.671)(0.730)(0.813) & - \\
   \hline
\end{tabular}}
\end{center}
\end{table}

\begin{table}[h]
\begin{center}
\caption{Simulations results obtained on 500 datasets with 400 families of size 5 with at least 1 case of a rare disease (prevalence around 1\%) and $\alpha_{1}$=0.1. Into brackets are respectively standard deviations, root mean square errors and the coverage probabilities.}
\scalebox{0.7}{
\begin{tabular}{r c c c}
\hline
 & Real Value & Retrospective likelihood & Naive method \\
 && Est (SD) (RMSE) (Cov Pr) & Est (SD) (RMSE) (Cov Pr)\\
\hline
 $\beta_0$ & 3.500 & 3.439(1.190)(1.216)(0.912) & 4.361(0.194)(0.883)(0.002) \\
 $\beta_1$ & 0.200 & 0.200(0.209)(0.213)(0.955) & 0.192(0.214)(0.216)(0.950) \\
 $\sigma_{GX}$ & 2.000 & 1.983(0.257)(0.278)(0.679) & 1.363(0.195)(0.683)(0.054) \\
 $\sigma_{\epsilon}$ & 1.414 & 1.552(0.119)(0.234)(0.644) & 2.356(0.058)(0.951)(0.000) \\
 $\alpha_0$ & -2.326 & -2.470(1.951)(2.169)(0.902) & - \\
 $\alpha_1$ & 0.100 & 0.074(0.174)(0.191)(0.932) & - \\
 $\sigma_{GY}$ & 1.732 & 1.553(0.574)(0.973)(0.765) & - \\
 $\sigma_{u}$ & 1.414 & 0.973(0.677)(0.745)(0.862) & - \\
   \hline
\end{tabular}}
\end{center}
\end{table}

\begin{table}[h]
\begin{center}
\caption{Simulations results obtained on 500 datasets with 400 families of size 5 with at least 1 case of a rare disease (prevalence around 1\%) and $\alpha_{1}$=0.5. Into brackets are respectively standard deviations, root mean square errors and the coverage probabilities.}
\scalebox{0.7}{
\begin{tabular}{r c c c}
\hline
 & Real Value & Retrospective likelihood & Naive method \\
 && Est (SD) (RMSE) (Cov Pr) & Est (SD) (RMSE) (Cov Pr)\\
\hline
 $\beta_0$ & 3.500 & 3.377(0.940)(0.927)(0.949) & 4.293(0.199)(0.816)(0.008) \\
 $\beta_1$ & 0.200 & 0.195(0.218)(0.218)(0.943) & 0.155(0.202)(0.218)(0.920) \\
 $\sigma_{GX}$ & 2.000 & 1.995(0.233)(0.233)(0.704) & 1.409(0.188)(0.641)(0.092) \\
 $\sigma_{\epsilon}$ & 1.414 & 1.544(0.115)(0.213)(0.674) & 2.329(0.059)(0.924)(0.000) \\
 $\alpha_0$ & -2.326 & -2.342(1.530)(1.536)(0.961) & - \\
 $\alpha_1$ & 0.500 & 0.412(0.156)(0.283)(0.877) & - \\
 $\sigma_{GY}$ & 1.732 & 1.578(0.556)(0.983)(0.787) & - \\
 $\sigma_{u}$ & 1.414 & 1.027(0.597)(0.698)(0.888) & - \\
 \hline
\end{tabular}}
\end{center}
\end{table}

\begin{table}[h]
\begin{center}
\caption{Simulations results obtained on 500 datasets with 400 families of size 5 with at least 2 cases of a common disease (prevalence around 5\%) and $\alpha_{1}$=0.1. Into brackets are respectively standard deviations, root mean square errors and the coverage probabilities.}
\scalebox{0.7}{
\begin{tabular}{r c c c}
\hline
 & Real Value & Retrospective likelihood & Naive method \\
 && Est (SD) (RMSE) (Cov Pr) & Est (SD) (RMSE) (Cov Pr)\\
\hline
$\beta_0$ & 3.500 & 3.397(1.463)(1.530)(0.901) & 4.835(0.180)(1.356)(0.000)\\
  $\beta_1$ & 0.200 & 0.202(0.204)(0.205)(0.939) & 0.188(0.203)(0.207)(0.940)\\
 $\sigma_{GX}$ & 2.000 & 1.919(0.467)(0.653)(0.650) & 1.053(0.362)(0.990)(0.060)\\
 $\sigma_{\epsilon}$ & 1.414 & 1.543(0.131)(0.238)(0.699) & 2.404(0.052)(0.998)(0.000)\\
 $\alpha_0$ & -1.644 & -1.755(1.801)(2.091)(0.874) & - \\
 $\alpha_1$ & 0.100 & 0.075(0.169)(0.222)(0.902) & - \\
$\sigma_{GY}$ & 1.732 & 1.470(0.669)(1.035)(0.837) & - \\
 $\sigma_{u}$ & 1.414 & 0.991(0.730)(0.765)(0.798) & -\\
 \hline
\end{tabular}}
\end{center}
\end{table}

\begin{table}[h]
\begin{center}
\caption{Simulations results obtained on 500 datasets with 400 families of size 5 with at least 2 cases of a common disease (prevalence around 5\%) and $\alpha_{1}$=0.5. Into brackets are respectively standard deviations, root mean square errors and the coverage probabilities.}
\scalebox{0.7}{
\begin{tabular}{r c c c}
\hline
 & Real Value & Retrospective likelihood & Naive method \\
 && Est (SD) (RMSE) (Cov Pr) & Est (SD) (RMSE) (Cov Pr)\\
\hline
 $\beta_0$ & 3.500 & 3.393(1.179)(1.156)(0.952) & 4.753(0.187)(0.1.266)(0.000) \\
 $\beta_1$ & 0.200 & 0.197(0.212)(0.215)(0.946) & 0.103(0.201)(0.221)(0.930) \\
 $\sigma_{GX}$ & 2.000 & 2.003(0.327)(0.513)(0.700) & 1.131(0.299)(0.912)(0.032) \\
 $\sigma_{\epsilon}$ & 1.414 & 1.523(0.114)(0.193)(0.702) & 2.376(0.053)(0.970)(0.000) \\
 $\alpha_0$ & -1.644 & -1.623(1.432)(1.663)(0.952) & - \\
 $\alpha_1$ & 0.500 & 0.409(0.170)(0.276)(0.926) & - \\
 $\sigma_{GY}$ & 1.732 & 1.530(0.570)(0.967)(0.864) & - \\
 $\sigma_{u}$ & 1.414 & 1.033(0.725)(0.763)(0.878) & - \\
 \hline
\end{tabular}}
\end{center}
\end{table}

\begin{table}[h]
\begin{center}
\caption{Simulations results obtained on 500 datasets with 400 families of size 5 with at least 1 case of a common disease (prevalence around 5\%) and $\alpha_{1}$=0.1. Into brackets are respectively standard deviations, root mean square errors and the coverage probabilities.}
\scalebox{0.7}{
\begin{tabular}{r c c c}
\hline
 & Real Value & Retrospective likelihood & Naive method \\
 && Est (SD) (RMSE) (Cov Pr) & Est (SD) (RMSE) (Cov Pr)\\
\hline
$\beta_0$ & 3.500 & 3.285(1.203)(1.288)(0.904) & 4.124(0.198)(0.653)(0.110) \\
  $\beta_1$ & 0.200 & 0.203(0.209)(0.209)(0.946) & 0.191(0.205)(0.217)(0.940) \\
 $\sigma_{GX}$ & 2.000 & 1.999(0.237)(0.479)(0.660) & 1.471(0.172)(0.586)(0.134) \\
 $\sigma_{\epsilon}$ & 1.414 & 1.559(0.116)(0.221)(0.672) & 2.291(0.061)(0.887)(0.000) \\
 $\alpha_0$ & -1.644 & -1.991(1.870)(2.450)(0.920) & - \\
 $\alpha_1$ & 0.100 & 0.085(0.173)(0.221)(0.916) & - \\
$\sigma_{GY}$ & 1.732 & 1.623(0.573)(1.033)(0.742) & - \\
 $\sigma_{u}$ & 1.414 & 0.960(0.738)(0.761)(0.894) & -  \\
 \hline
\end{tabular}}
\end{center}
\end{table}

\begin{table}[h]
\begin{center}
\caption{Simulations results obtained on 500 datasets with 400 families of size 5 with at least 1 case of a common disease (prevalence around 5\%) and $\alpha_{1}$=0.5. Into brackets are respectively standard deviations, root mean square errors and the coverage probabilities.}
\scalebox{0.7}{
\begin{tabular}{r c c c}
\hline
 & Real Value & Retrospective likelihood & Naive method \\
 && Est (SD) (RMSE) (Cov Pr) & Est (SD) (RMSE) (Cov Pr)\\
\hline
$\beta_0$ & 3.500 & 3.156(0.939)(0.973)(0.946) & 4.066(0.201)(0.599)(0.200) \\
  $\beta_1$ & 0.200 & 0.194(0.222)(0.226)(0.944) & 0.158(0.204)(0.234)(0.920) \\
 $\sigma_{GX}$ & 2.000 & 2.035(0.245)(0.254)(0.712) & 1.520(0.163)(0.540)(0.176) \\
 $\sigma_{\epsilon}$ & 1.414 & 1.543(0.124)(0.219)(0.654) & 2.263(0.063)(0.861)(0.000) \\
 $\alpha_0$ & -1.644 & -1.982(1.518)(1.612)(0.960) & - \\
 $\alpha_1$ & 0.500 & 0.430(0.304)(0.574)(0.898) & - \\
$\sigma_{GY}$ & 1.732 & 1.680(0.574)(1.090)(0.772) & - \\
 $\sigma_{u}$ & 1.414 & 0.986(0.666)(0.718)(0.894) & -  \\
 \hline
\end{tabular}}
\end{center}
\end{table}

\begin{table}[h]
\begin{center}
\caption{Estimates with standard deviations and RMSE of the heritability of the secondary phenotype are given for a common disease (prevalence $\approx$ 5\%), for the four ascertainment mechanisms and two values of $\alpha_{1}$.}
\scalebox{0.7}{
\begin{tabular*}{0.85\textwidth}{ l  c  c c }
\hline
Ascertainment & $\alpha_{1}$ & Retrospective likelihood & Naive method\\
\hline

1.At least 2 cases & & & \\[2pt]
& 0.10 & 0.49(0.08)(0.21) & 0.17(0.08)(0.34)\\
& 0.50 & 0.50(0.08)(0.18) & 0.19(0.08)(0.32)\\

2.At least 1 case & & & \\[2pt]
& 0.10 & 0.50(0.08)(0.17) & 0.29(0.08)(0.22)\\
& 0.50 & 0.52(0.08)(0.16) & 0.31(0.08)(0.20)\\
\hline
\end{tabular*}}
\end{center}
\end{table}

\begin{table}[h]
\begin{center}
\caption{Estimates with standard deviations and RMSE of the heritability of the secondary phenotype are given for a rare disease (prevalence $\approx$ 1\%), for the four ascertainment mechanisms and two values of $\alpha_{1}$.}
\scalebox{0.7}{
\begin{tabular*}{0.85\textwidth}{ l  c  c c }
\hline
Ascertainment & $\alpha_{1}$ & Retrospective likelihood & Naive method\\
\hline

1.At least 2 cases & & & \\[2pt]
& 0.10 & 0.48(0.07)(0.22) & 0.13(0.07)(0.37)\\
& 0.50 & 0.48(0.07)(0.22) & 0.14(0.07)(0.36)\\

2.At least 1 case & & & \\[2pt]
& 0.10 & 0.50(0.08)(0.17) & 0.25(0.08)(0.25)\\
& 0.50 & 0.50(0.08)(0.17) & 0.27(0.08)(0.24)\\
\hline
\end{tabular*}}
\end{center}
\end{table}

\begin{figure}[h]
\begin{center}
\includegraphics[scale=0.3]{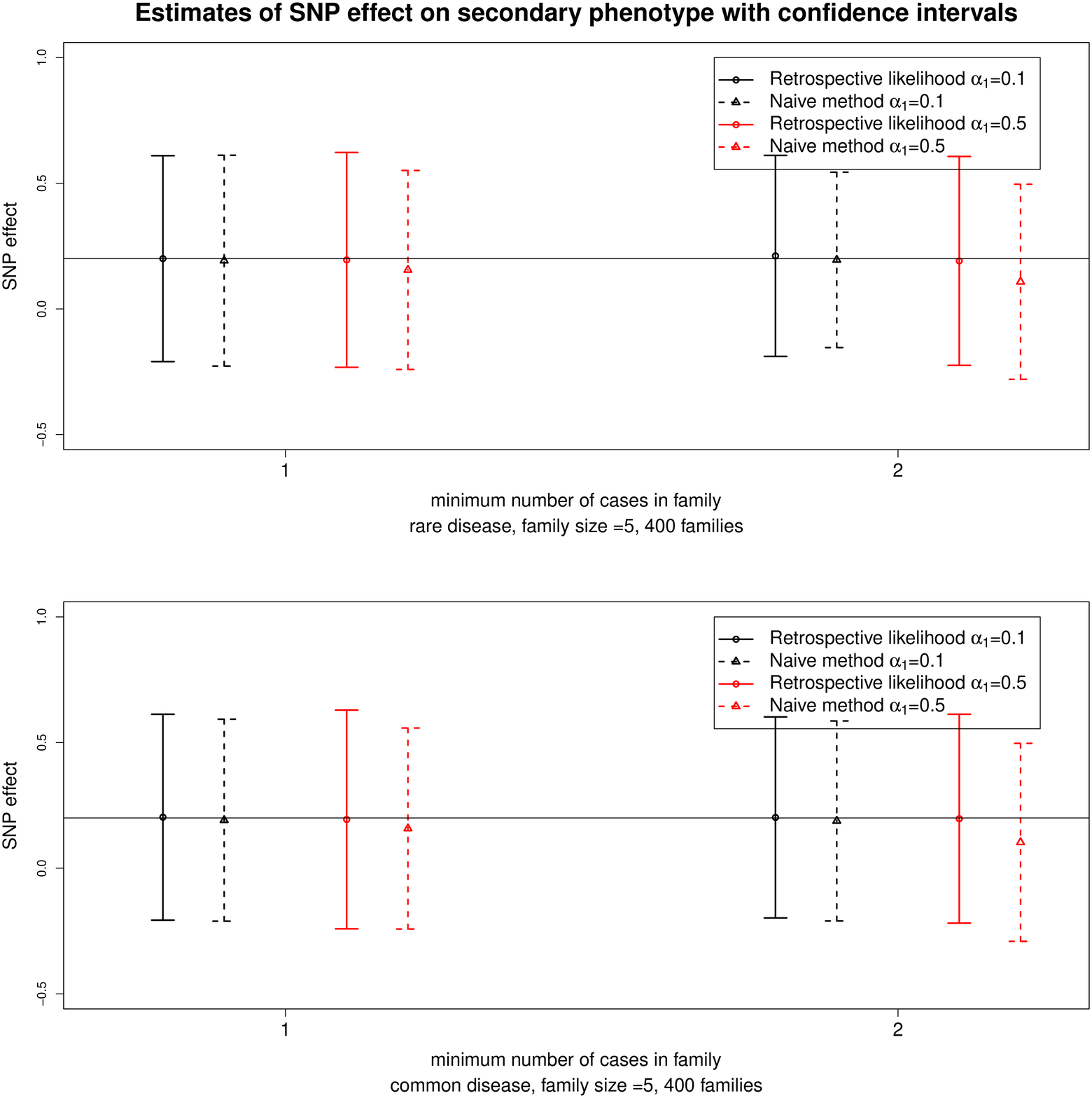}
\caption{Estimates and 95\% confidence intervals for the SNP effect on the secondary phenotype for the retrospective likelihood approach and the naive method. Results are obtained from 500 simulated datasets of 100 families for 2 ascertainment schedules. The top and bottom panel correspond to a rare or common primary phenotype with a prevalence around 1\% and 5\% respectively. In black and red are represented results for small ($\alpha_1$=0.5) and large ($\alpha_1$=1) effect sizes of the SNP on the primary phenotype respectively. The horizontal line corresponds to the true SNP effect on the secondary phenotype.}
\label{Fig2}
\end{center}
\end{figure}

\clearpage

\begin{table}[h]
\begin{center}
\caption{Type I errors rates for testing for association between a marker and a secondary phenotype for four scenarios. Two ascertainment processes, namely at least one and at least two cases are considered. Two values for the association between the SNP and the primary phenotype are used, namely $\alpha_{1}$ = 0.1 and $\alpha_{1}$ = 0.5. We simulated for each scenario 10000 datasets.}
\scalebox{0.7}{
\begin{tabular*}{0.8\textwidth}{ l r c c}
\hline
 & nominal level ($\alpha$) & Retrospective likelihood &  Naive method \\
\hline
At least 2 cases  &  & &\\
$\alpha_{1}$ =0.1 & & &\\
& 0.05  & 0.0509  & 0.0580 \\
 & 0.01 & 0.0118  & 0.0152 \\
 & 0.001 & 0.0017  & 0.0025  \\
$\alpha_{1}$ =0.5 & & &\\
& 0.05 & 0.0505  & 0.0878  \\
& 0.01 & 0.0113  & 0.0222  \\
& 0.001 & 0.0013 & 0.0043  \\
At least 1 case  &  & &\\
$\alpha_{1}$ =0.1  & & &\\
& 0.05 & 0.0524& 0.0514   \\
 &0.01 & 0.0102 & 0.0098  \\
 & 0.001 & 0.0018 & 0.0014 \\
$\alpha_{1}$ =0.5  & & &\\
& 0.05 & 0.0522 & 0.0558 \\
& 0.01 & 0.0098 & 0.0097 \\
& 0.001 & 0.0009 & 0.0016 \\
\cline{1-4}
\end{tabular*}}
\end{center}
\end{table}

\clearpage

\begin{table}[h]
\begin{center}
\caption{Simulations results obtained on 500 datasets with 400 families of size 5 with at least 2 cases of a rare disease (prevalence around 1\%) and $\alpha_{1}$=0.1. Into brackets are respectively standard deviations, root mean square errors and the coverage probabilities.}
\scalebox{0.7}{
\begin{tabular}{r c c c}
\hline
 & Real Value & Retrospective likelihood  \\
 && Est (SD) (RMSE) (Cov Pr) \\
\hline
 $\beta_0$ & 3.500 & 3.586(0.897)(0.910)(0.905)\\
 $\beta_1$ & 0.200 & 0.199(0.103)(0.104)(0.948)\\
 $\sigma_{GX}$ & 2.000 & 1.887(0.155)(0.296)(0.623)\\
 $\sigma_{\epsilon}$ & 1.414 & 1.571(0.050)(0.178)(0.257)\\
 $\alpha_0$ & -2.326 & -1.366(0.838)(1.342)(0.659)\\
 $\alpha_1$ & 0.100 & 0.053(0.094)(0.250)(0.901)  \\
 $\sigma_{GY}$ & 1.732 & 1.019(0.250)(0.764)(0.129) \\
 $\sigma_{u}$ & 1.414 & 0.988(0.644)(0.526)(0.948)  \\
\hline
\end{tabular}}
\end{center}
\end{table}

\begin{table}[h]
\begin{center}
\caption{Simulations results obtained on 500 datasets with 400 families of size 5 with at least 2 cases of a rare disease (prevalence around 1\%) and $\alpha_{1}$=0.5. Into brackets are respectively standard deviations, root mean square errors and the coverage probabilities.}
\scalebox{0.7}{
\begin{tabular}{r c c }
\hline
 & Real Value & Retrospective likelihood \\
 && Est (SD) (RMSE) (Cov Pr) \\
\hline
 $\beta_0$ & 3.500 & 3.720(0.583)(0.625)(0.905)\\
 $\beta_1$ & 0.200 & 0.197(0.106)(0.110)(0.945) \\
 $\sigma_{GX}$ & 2.000 & 1.909(0.139)(0.271)(0.625)\\
 $\sigma_{\epsilon}$ & 1.414 & 1.560(0.047)(0.162)(0.245)\\
 $\alpha_0$ & -2.326 & -1.161(0.528)(1.245)(0.311)\\
 $\alpha_1$ & 0.500 & 0.287(0.221)(0.232)(0.913)\\
 $\sigma_{GY}$ & 1.732 & 1.052(0.221)(0.715)(0.094\\
 $\sigma_{u}$ & 1.414 & 1.009(0.501)(0.601)(0.975)\\
\hline
\end{tabular}}
\end{center}
\end{table}

\begin{table}[h]
\begin{center}
\caption{Simulations results obtained on 500 datasets with 400 families of size 5 with at least 1 case of a rare disease (prevalence around 1\%) and $\alpha_{1}$=0.1. Into brackets are respectively standard deviations, root mean square errors and the coverage probabilities.}
\scalebox{0.7}{
\begin{tabular}{r c c c}
\hline
 & Real Value & Retrospective likelihood \\
 && Est (SD) (RMSE) (Cov Pr) \\
\hline
 $\beta_0$ & 3.500 & 3.630(0.726)(0.730)(0.909)\\
 $\beta_1$ & 0.200 & 0.200(0.104)(0.107)(0.961)\\
 $\sigma_{GX}$ & 2.000 & 1.887(0.133)(0.266)(0.623)\\
 $\sigma_{\epsilon}$ & 1.414 & 1.572(0.050)(0.176)(0.248)\\
 $\alpha_0$ & -2.326 & -1.377(0.784)(1.339)(0.657)\\
 $\alpha_1$ & 0.100 & 0.056(0.090)(0.231)(0.900)\\
 $\sigma_{GY}$ & 1.732 & 1.010(0.231)(0.771)(0.116)\\
 $\sigma_{u}$ & 1.414 & 0.998(0.495)(0.514)(0.761)\\
\hline
\end{tabular}}
\end{center}
\end{table}

\begin{table}[h]
\begin{center}
\caption{Simulations results obtained on 500 datasets with 400 families of size 5 with at least 1 case of a rare disease (prevalence around 1\%) and $\alpha_{1}$=0.5. Into brackets are respectively standard deviations, root mean square errors and the coverage probabilities.}
\scalebox{0.7}{
\begin{tabular}{r c c }
\hline
 & Real Value & Retrospective likelihood \\
 && Est (SD) (RMSE) (Cov Pr) \\
\hline
 $\beta_0$ & 3.500 & 3.440(0.366)(0.467)(0.996)\\
 $\beta_1$ & 0.200 & 0.199(0.107)(0.111)(0.960)\\
 $\sigma_{GX}$ & 2.000 & 1.892(0.108)(0.231)(0.670) \\
 $\sigma_{\epsilon}$ & 1.414 & 1.4(0.045)(0.171)(0.284)\\
 $\alpha_0$ & -2.326 & -1.453(0.472)(0.940)(0.446)\\
 $\alpha_1$ & 0.500 & 0.295(0.190)(0.222)(0.948) \\
 $\sigma_{GY}$ & 1.732 & 1.049(0.190)(0.711)(0.074)\\
 $\sigma_{u}$ & 1.414 & 1.000(0.572)(0.494)(0.680) \\
 \hline
\end{tabular}}
\end{center}
\end{table}

\begin{table}[h]
\begin{center}
\caption{Simulations results obtained on 500 datasets with 400 families of size 5 with at least 2 cases of a common disease (prevalence around 5\%) and $\alpha_{1}$=0.1. Into brackets are respectively standard deviations, root mean square errors and the coverage probabilities.}
\scalebox{0.7}{
\begin{tabular}{r c c }
\hline
 & Real Value & Retrospective likelihood  \\
 && Est (SD) (RMSE) (Cov Pr) \\
\hline
 $\beta_0$ & 3.500 & 3.500(0.822)(0.871)(0.961)\\
 $\beta_1$ & 0.200 & 0.193(0.105)(0.105)(0.951)\\
 $\sigma_{GX}$ & 2.000 & 1.884(0.140)(0.278)(0.597)\\
 $\sigma_{\epsilon}$ & 1.414 & 1.572(0.047)(0.175)(0.236)\\
 $\alpha_0$ & -1.644 & -1.021(0.804)(1.050)(0.813)\\
 $\alpha_1$ & 0.100 & 0.051(0.098)(0.230)(1.000)\\
 $\sigma_{GY}$ & 1.732 & 1.040(0.230)(0.736)(0.131)\\
 $\sigma_{u}$ & 1.414 & 0.987(0.517)(0.602)(0.931)\\
\hline
\end{tabular}}
\end{center}
\end{table}

\begin{table}[h]
\begin{center}
\caption{Simulations results obtained on 500 datasets with 400 families of size 5 with at least 2 cases of a common disease (prevalence around 5\%) and $\alpha_{1}$=0.5. Into brackets are respectively standard deviations, root mean square errors and the coverage probabilities.}
\scalebox{0.7}{
\begin{tabular}{r c c }
\hline
 & Real Value & Retrospective likelihood  \\
 && Est (SD) (RMSE) (Cov Pr) \\
\hline
 $\beta_0$ & 3.500 & 3.501(0.589)(0.449)(0.996)\\
 $\beta_1$ & 0.200 & 0.186(0.107)(0.108)(0.940)\\
 $\sigma_{GX}$ & 2.000 & 1.886(0.125)(0.246)(0.606)\\
 $\sigma_{\epsilon}$ & 1.414 & 1.568(0.044)(0.168)(0.174)\\
 $\alpha_0$ & -1.644 & -0.995(0.508)(0.767)(0.650)\\
 $\alpha_1$ & 0.500 & 0.289(0.207)(0.232)(0.904)\\
 $\sigma_{GY}$ & 1.732 & 1.054(0.207)(0.709)(0.088)\\
 $\sigma_{u}$ & 1.414 & 1.015(0.480)(0.517)(0.910)\\
\hline
\end{tabular}}
\end{center}
\end{table}

\begin{table}[h]
\begin{center}
\caption{Simulations results obtained on 500 datasets with 400 families of size 5 with at least 1 case of a common disease (prevalence around 5\%) and $\alpha_{1}$=0.1. Into brackets are respectively standard deviations, root mean square errors and the coverage probabilities.}
\scalebox{0.7}{
\begin{tabular}{r c c }
\hline
 & Real Value & Retrospective likelihood  \\
 && Est (SD) (RMSE) (Cov Pr) \\
\hline
 $\beta_0$ & 3.500 & 3.509(0.716)(0.720)(0.921)\\
 $\beta_1$ & 0.200 & 0.197(0.106)(0.108)(0.949)\\
 $\sigma_{GX}$ & 2.000 & 1.884(0.117)(0.254)(0.587)\\
 $\sigma_{\epsilon}$ & 1.414 & 1.576(0.046)(0.178)(0.297)\\
 $\alpha_0$ & -1.644 & -1.069(0.730)(1.078)(0.823)\\
 $\alpha_1$ & 0.100 & 0.055(0.094)(0.204)(0.905)\\
 $\sigma_{GY}$ & 1.732 & 1.036(0.204)(0.741)(0.109)\\
 $\sigma_{u}$ & 1.414 & 0.988(0.499)(0.512)(0.724)\\
\hline
\end{tabular}}
\end{center}
\end{table}

\begin{table}[h]
\begin{center}
\caption{Simulations results obtained on 500 datasets with 400 families of size 5 with at least 1 case of a common disease (prevalence around 5\%) and $\alpha_{1}$=0.5. Into brackets are respectively standard deviations, root mean square errors and the coverage probabilities.}
\scalebox{0.7}{
\begin{tabular}{r c c }
\hline
 & Real Value & Retrospective likelihood  \\
 && Est (SD) (RMSE) (Cov Pr) \\
\hline
 $\beta_0$ & 3.500 & 3.254(0.421)(0.449)(0.982)\\
 $\beta_1$ & 0.200 & 0.196(0.107)(0.108)(0.942)\\
 $\sigma_{GX}$ & 2.000 & 1.868(0.099)(0.226)(0.652)\\
 $\sigma_{\epsilon}$ & 1.414 & 1.575(0.043)(0.175)(0.260)\\
 $\alpha_0$ & -1.644 & -1.249(0.440)(0.540)(0.936)\\
 $\alpha_1$ & 0.500 & 0.302(0.181)(0.217)(0.948)\\
 $\sigma_{GY}$ & 1.732 & 1.044(0.181)(0.711)(0.046)\\
 $\sigma_{u}$ & 1.414 & 1.022(0.383)(0.450)(0.770)\\
\hline
\end{tabular}}
\end{center}
\end{table}

\begin{table}[h]
\begin{center}
\caption{Estimates with standard deviations and RMSE of the heritability of the secondary phenotype are given for a common disease (prevalence $\approx$ 5\%), for the four ascertainment mechanisms and two values of $\alpha_{1}$.}
\scalebox{0.7}{
\begin{tabular*}{0.85\textwidth}{ l  c  c c }
\hline
Ascertainment & $\alpha_{1}$ & Retrospective likelihood & Naive method\\
\hline

1.At least 2 cases & & & \\[2pt]
& 0.100 & 0.499(0.102)(0.102) & 0.17(0.08)(0.34)\\
& 0.500 & 0.498(0.089)(0.089) & 0.19(0.08)(0.32)\\

2.At least 1 case & & & \\[2pt]
& 0.100 & 0.499(0.089)(0.089) & 0.29(0.08)(0.22)\\
& 0.500 & 0.492(0.076)(0.077) & 0.31(0.08)(0.20)\\
\hline
\end{tabular*}}
\end{center}
\end{table}

\begin{table}[h]
\begin{center}
\caption{Estimates with standard deviations and RMSE of the heritability of the secondary phenotype are given for a rare disease (prevalence $\approx$ 1\%), for the four ascertainment mechanisms and two values of $\alpha_{1}$.}
\scalebox{0.7}{
\begin{tabular*}{0.85\textwidth}{ l  c  c }
\hline
Ascertainment & $\alpha_{1}$ & Retrospective likelihood\\
\hline

1.At least 2 cases & & \\[2pt]
& 0.100 & 0.499(0.107)(0.107))\\
& 0.500 & 0.505(0.104)(0.104)\\

2.At least 1 case & & \\[2pt]
& 0.100 & 0.499(0.092)(0.092))\\
& 0.500 & 0.501(0.084)(0.084)\\
\hline
\end{tabular*}}
\end{center}
\end{table}

\clearpage

\subsubsection{Simulation study for a polygenic score}

\begin{table}[h]
\begin{center}
\caption{Simulations results obtained on 500 datasets with 400 families of size 5 with at least 2 cases of a rare disease (prevalence around 1\%) and $\alpha_{1}$=0.1. Into brackets are respectively standard deviations, root mean square errors and the coverage probabilities.}
\scalebox{0.7}{
\begin{tabular}{r c c c}
\hline
 & Real Value & Retrospective likelihood & Naive method \\
 && Est (SD) (RMSE) (Cov Pr) & Est (SD) (RMSE) (Cov Pr)\\
\hline
 $\beta_0$ & 3.500 & 3.483(0.899)(0.907)(0.899) & 5.221(0.067)(1.723)(0.000) \\
 $\beta_1$ & 0.200 & 0.193(0.065)(0.073)(0.932) & 0.175(0.062)(0.070)(0.922) \\
 $\sigma_{GX}$ & 2.000 & 2.013(0.190)(0.357)(0.657) & 0.990(0.146)(1.018131)(0.000)\\
 $\sigma_{\epsilon}$ & 1.414 & 1.529(0.055)(0.140)(0.559) & 2.467(0.025)(1.054)(0.000) \\
 $\alpha_0$ & -2.326 & -1.932(0.921)(1.013)(0.839) &- \\
 $\alpha_1$ & 0.100 & 0.076(0.054)(0.069)(0.814) & - \\
 $\sigma_{GY}$ & 1.732 & 1.398(0.328)(0.548)(0.793) & - \\
 $\sigma_{u}$ & 1.414 & 1.170(0.778)(0.869)(0.920) & - \\
\hline
\end{tabular}}
\end{center}
\end{table}

\begin{table}[h]
\begin{center}
\caption{Simulations results obtained on 500 datasets with 400 families of size 5 with at least 2 cases of a rare disease (prevalence around 1\%) and $\alpha_{1}$=0.5. Into brackets are respectively standard deviations, root mean square errors and the coverage probabilities.}
\scalebox{0.7}{
\begin{tabular}{r c c c}
\hline
 & Real Value & Retrospective likelihood & Naive method \\
 && Est (SD) (RMSE) (Cov Pr) & Est (SD) (RMSE) (Cov Pr)\\
\hline
 $\beta_0$ & 3.500 & 3.962(0.541)(0.618)(0.834) & 5.178(0.069)(1.679)(0.000) \\
 $\beta_1$ & 0.200 & 0.193(0.074)(0.079)(0.940) & 0.096(0.063)(0.122)(0.610) \\
 $\sigma_{GX}$ & 2.000 & 2.084(0.165)(0.346)(0.664) & 1.037(0.135)(0.972)(0.000) \\
 $\sigma_{\epsilon}$ & 1.414 & 1.518(0.053)(0.130)(0.592) & 2.452(0.025)(1.039)(0.000) \\
 $\alpha_0$ & -2.326 & -1.871(0.522)(1.0314)(0.758) & - \\
 $\alpha_1$ & 0.500 & 0.391(0.074)(0.134)(0.772) & - \\
 $\sigma_{GY}$ & 1.732 & 1.471(0.273)(0.475)(0.796) & - \\
 $\sigma_{u}$ & 1.414 & 1.176(0.670)(0.744)(0.924) & - \\
   \hline
\end{tabular}}
\end{center}
\end{table}

\begin{table}[h]
\begin{center}
\caption{Simulations results obtained on 500 datasets with 400 families of size 5 with at least 1 case of a rare disease (prevalence around 1\%) and $\alpha_{1}$=0.1. Into brackets are respectively standard deviations, root mean square errors and the coverage probabilities.}
\scalebox{0.7}{
\begin{tabular}{r c c c}
\hline
 & Real Value & Retrospective likelihood & Naive method \\
 && Est (SD) (RMSE) (Cov Pr) & Est (SD) (RMSE) (Cov Pr)\\
\hline
 $\beta_0$ & 3.500 & 3.338(0.573)(0.655)(0.862) & 4.370(0.076)(0.850)(0.000) \\
 $\beta_1$ & 0.200 & 0.190(0.067)(0.067)(0.954) & 0.189(0.066)(0.066)(0.948) \\
 $\sigma_{GX}$ & 2.000 & 1.925(0.163)(0.364)(0.645) & 1.386(0.088)(0.626)(0.000) \\
 $\sigma_{\epsilon}$ & 1.414 & 1.460(0.057)(0.171)(0.676) & 2.364(0.028)(0.952)(0.000) \\
 $\alpha_0$ & -2.326 & -2.219(0.774)(0.922)(0.889) & - \\
 $\alpha_1$ & 0.100 & 0.073(0.058)(0.071)(0.829) & - \\
 $\sigma_{GY}$ & 1.732 & 1.359(0.307)(0.583)(0.611) & - \\
 $\sigma_{u}$ & 1.414 & 1.186(0.623)(0.714)(0.926) & - \\
   \hline
\end{tabular}}
\end{center}
\end{table}

\begin{table}[h]
\begin{center}
\caption{Simulations results obtained on 500 datasets with 400 families of size 5 with at least 1 case of a rare disease (prevalence around 1\%) and $\alpha_{1}$=0.5. Into brackets are respectively standard deviations, root mean square errors and the coverage probabilities.}
\scalebox{0.7}{
\begin{tabular}{r c c c}
\hline
 & Real Value & Retrospective likelihood & Naive method \\
 && Est (SD) (RMSE) (Cov Pr) & Est (SD) (RMSE) (Cov Pr)\\
\hline
 $\beta_0$ & 3.500 & 3.542(0.361)(0.411)(0.963) & 4.346(0.076)(0.850)(0.000) \\
 $\beta_1$ & 0.200 & 0.197(0.074)(0.078)(0.944) & 0.157(0.066)(0.079)(0.904) \\
 $\sigma_{GX}$ & 2.000 & 2.018(0.127)(0.278)(0.663) & 1.412(0.186)(0.600)(0.000) \\
 $\sigma_{\epsilon}$ & 1.414 & 1.538(0.052)(0.147)(0.582) & 2.352(0.029)(0.940)(0.000) \\
 $\alpha_0$ & -2.326 & -1.876(0.425)(631)(0.797) & - \\
 $\alpha_1$ & 0.500 & 0.401(0.074)(0.128)(0.838) & - \\
 $\sigma_{GY}$ & 1.732 & 1.435(0.240)(0.473)(0.619) & - \\
 $\sigma_{u}$ & 1.414 & 1.209(0.466)(0.574)(0.948) & - \\
 \hline
\end{tabular}}
\end{center}
\end{table}

\begin{table}[h]
\begin{center}
\caption{Simulations results obtained on 500 datasets with 400 families of size 5 with at least 2 cases of a common disease (prevalence around 5\%) and $\alpha_{1}$=0.1. Into brackets are respectively standard deviations, root mean square errors and the coverage probabilities.}
\scalebox{0.7}{
\begin{tabular}{r c c c}
\hline
 & Real Value & Retrospective likelihood & Naive method \\
 && Est (SD) (RMSE) (Cov Pr) & Est (SD) (RMSE) (Cov Pr)\\
\hline
$\beta_0$ & 3.500 & 3.370(0.854)(0.854)(0.933) & 4.854(0.069)(1.356)(0.000)\\
  $\beta_1$ & 0.200 & 0.192(0.067)(0.067)(0.952) & 0.177(0.063)(0.063)(0.948)\\
 $\sigma_{GX}$ & 2.000 & 2.018(0.127)(0.278)(0.663) & 1.113(0.118)(0.896)(0.000)\\
 $\sigma_{\epsilon}$ & 1.414 & 1.538(0.052)(0.147)(0.682) & 2.405(0.026)(0.992)(0.000)\\
 $\alpha_0$ & -1.644 & -1.470(0.875)(0.897)(0.897) & - \\
 $\alpha_1$ & 0.100 & 0.073(0.057)(0.068)(0.879) & - \\
$\sigma_{GY}$ & 1.732 & 1.477(0.295)(0.500)(0.772) & - \\
 $\sigma_{u}$ & 1.414 & 1.099(0.997)(1.139)(0.945) & - \\
\hline
\end{tabular}}
\end{center}
\end{table}

\begin{table}[h]
\begin{center}
\caption{Simulations results obtained on 500 datasets with 400 families of size 5 with at least 2 cases of a common disease (prevalence around 5\%) and $\alpha_{1}$=0.5. Into brackets are respectively standard deviations, root mean square errors and the coverage probabilities.}
\scalebox{0.7}{
\begin{tabular}{r c c c}
\hline
 & Real Value & Retrospective likelihood & Naive method \\
 && Est (SD) (RMSE) (Cov Pr) & Est (SD) (RMSE) (Cov Pr)\\
\hline
 $\beta_0$ & 3.500 & 3.653(0.502)(0.593)(0.932) & 4.823(0.070)(1.325)(0.000) \\
 $\beta_1$ & 0.200 & 0.192(0.073)(0.075)(0.954) & 0.103(0.064)(0.115)(0.656) \\
 $\sigma_{GX}$ & 2.000 & 2.055(0.149)(0.309)(0.664) & 1.148(0.113)(0.861)(0.000) \\
 $\sigma_{\epsilon}$ & 1.414 & 1.530(0.051)(0.140)(0.614) & 2.395(0.026)(0.982)(0.000) \\
 $\alpha_0$ & -1.644 & -1.367(0.495)(0.599)(0.848) & - \\
 $\alpha_1$ & 0.500 & 0.394(0.072)(0.131)(0.868) & - \\
 $\sigma_{GY}$ & 1.732 & 1.462(0.255)(0.456)(0.762) & - \\
 $\sigma_{u}$ & 1.414 & 1.172(0.645)(0.733)(0.954) & -\\
\hline
\end{tabular}}
\end{center}
\end{table}

\begin{table}[h]
\begin{center}
\caption{Simulations results obtained on 500 datasets with 400 families of size 5 with at least 1 case of a common disease (prevalence around 5\%) and $\alpha_{1}$=0.1. Into brackets are respectively standard deviations, root mean square errors and the coverage probabilities.}
\scalebox{0.7}{
\begin{tabular}{r c c c}
\hline & Real Value & Retrospective likelihood & Naive method \\
 && Est (SD) (RMSE) (Cov Pr) & Est (SD) (RMSE) (Cov Pr)\\
\hline
$\beta_0$ & 3.500 & 3.088(0.524)(0.701)(0.896) & 4.134(0.077)(0.639)(0.000) \\
  $\beta_1$ & 0.200 & 0.196(0.066)(0.0.066)(0.960) & 0.193(0.066)(0.071)(0.962) \\
 $\sigma_{GX}$ & 2.000 & 1.936(0.144)(0.364)(0.694) & 1.482(0.080)(0.531)(0.000) \\
 $\sigma_{\epsilon}$ & 1.414 & 1.568(0.052)(0.218)(0.607) & 2.296(0.030)(0.884)(0.000) \\
 $\alpha_0$ & -1.644 & -1.955(0.669)(0.941)(0.953) & - \\
 $\alpha_1$ & 0.100 & 0.077(0.056)(0.065)(0.847) & - \\
$\sigma_{GY}$ & 1.732 & 1.444(0.285)(0.410)(0.687) & - \\
 $\sigma_{u}$ & 1.414 & 1.244(0.321)(0.415)(0.937) & -\\
   \hline
\end{tabular}}
\end{center}
\end{table}

\begin{table}[h]
\begin{center}
\caption{Simulations results obtained on 500 datasets with 400 families of size 5 with at least 1 case of a common disease (prevalence around 5\%) and $\alpha_{1}$=0.5. Into brackets are respectively standard deviations, root mean square errors and the coverage probabilities.}
\scalebox{0.7}{
\begin{tabular}{r c c c}
\hline
 & Real Value & Retrospective likelihood & Naive method \\
 && Est (SD) (RMSE) (Cov Pr) & Est (SD) (RMSE) (Cov Pr)\\
\hline
$\beta_0$ & 3.500 & 3.278(0.341)(0.408)(0.958) & 4.118(0.077)(0.623)(0.000) \\
  $\beta_1$ & 0.200 & 0.187(0.072)(0.078)(0.953) & 0.159(0.067)(0.074)(0.928) \\
 $\sigma_{GX}$ & 2.000 & 1.952(0.118)(0.285)(0.610) & 1.508(0.078)(0.506)(0.004) \\
 $\sigma_{\epsilon}$ & 1.414 & 1.552(0.049)(0.161)(0.607) & 2.286(0.030)(0.874)(0.000) \\
 $\alpha_0$ & -1.644 & -1.648(0.395)(0.567)(0.988) & - \\
 $\alpha_1$ & 0.500 & 0.392(0.071)(0.140)(0.798) & - \\
$\sigma_{GY}$ & 1.732 & 1.373(0.230)(0.504)(0.660) & - \\
 $\sigma_{u}$ & 1.414 & 1.223(0.328)(0.337)(0.943) & - \\ \hline
\end{tabular}}
\end{center}
\end{table}

\begin{table}[h]
\begin{center}
\caption{Estimates with standard deviations and RMSE of the heritability of the secondary phenotype are given for a rare disease (prevalence $\approx$ 1\%), for the four ascertainment mechanisms and two values of $\alpha_{1}$.}
\scalebox{0.7}{
\begin{tabular*}{0.85\textwidth}{ l  c  c c }
\hline
Ascertainment & $\alpha_{1}$ & Retrospective likelihood & Naive method\\
\hline

1.At least 2 cases & & & \\[2pt]
& 0.10 & 0.50(0.03)(0.13) & 0.14(0.03)(0.36)\\
& 0.50 & 0.52(0.03)0.12) & 0.15(0.03)(0.34)\\

2.At least 1 case & & & \\[2pt]
& 0.10 & 0.48(0.04)(0.12) & 0.25(0.03)(0.24)\\
& 0.50 & 0.50(0.04)(0.10) & 0.26(0.04)(0.23)\\
\hline
\end{tabular*}}
\end{center}
\end{table}

\clearpage

\subsection{Results analysis Leiden Longevity Study on triglyceride levels}

\begin{table}[h]
\begin{center}
\caption{Leiden Longevity Study: Estimates of the association between the 41 selected SNPS and triglyceride levels for womend and for three different approaches. The retrospective likelihood approach with same variance assumed for the shared random effect, with different variances, and the naive approach. Are also presented the absolute difference between the estimates of the two last approaches with the first one. Into brackets are the standard errors.}
\scalebox{0.7}{
\begin{tabular}{rccccc}
\hline
SNPs & Constrained approach & \multicolumn{2}{c}{Not Constrained approach}& \multicolumn{2}{c}{Naive Likelihood}  \\
 & Estimates (SE) & Estimates (SE) & Difference & Estimates (SE)& Difference\\
\hline
 rs3863318\_A & .0242(.0205) & .0245(.0205) & .0003 & .0209(.0205) & .0033 \\
  rs2512139\_A & .0000(.0215) & .0000(.0345) & .0000 & -.0031(.0214) & .0031 \\
  rs7103514\_G & .0101(.0365) & .0101(.0275) & .0000 & .0104(.0367) & .0003 \\
  rs2512158\_A & -.0428(.0206) & -.0430(.0208)  & .0002 & -.0423(.0207) & .0005 \\
  rs11216648\_G & .0271(.0206) & .0272(.0206) & .0001 & .0276(.0207) & .0005 \\
  rs4936414\_G & -.0433(.0212) & -.430 (.0212) & .0003 &  -.0429(.0213) & .0004  \\
  rs4252287\_A & .0314(.0345) & .0313(.0385) & .0001 & .0303(.0347) & .0011 \\
  rs947889\_G & -.0239(.0202) & -.0238(.0199) & .0001 & -.0241(.0203) & .0002 \\
  rs2512154\_A & .0299(.0307) & .0298(.0215) & .0001 & .0295(.0308) & .0004 \\
  rs4936417\_G & .0104(.0223) & .0104(.0233) & .0000 & .0118(.0223) & .0014 \\
  rs652107\_G & -.0078(.0275) & -.0075(.206) & .0003 & -.0081(.0276) & .0003 \\
  rs12576767\_A & -.0329(.0432) & -.320(.0367) & .0009 & -.0299(.0433) & .0030 \\
  rs1786186\_A & -.0131(.0287) & -.0132(.0216) & .0001 & -.0094(.0287) & .0037 \\
  rs3825050\_A & -.0085(.0219) & -.0083(.0205) & .0002 & -.0112(.0219) & .0027 \\
  rs10160375\_G & -.0071(.0214) & -.0072(.0203) & .0001 & -.0092(.0215) & .0021 \\
  rs689264\_A & .0108(.0368) & .0107(.0431) & .0001 & .0116(.0370) & .0008 \\
  rs948461\_A & -.0075(.0216) & -.0072(.205) & .0003 & -.0067(.0217) & 0008 \\
  rs2276123\_A & -.0064(.0218) & .0070(.0243) & .0006  & -.0090(.0218) & .0026 \\
  rs948466\_A & .0116(.0385) & .0069(.0509) & .0047 & .0070(.0387) & .0046 \\
  rs881122\_A & .0149(.0243) & .0149(.0217) & .0000 & .0169(.0244) & .0020 \\
  rs2276129\_A & -.0012(.0217) & -.0008(.0217) & .0004 & -.0028(.0218) & .0016 \\
  rs2155857\_A & .0323(.0509) & .0320(.0200) & .0003 & .0344(.0511) & .0021 \\
  rs1894177\_A & -.0115(.0200) & -.0130(.0456) & .0015 & -.0118(.0201) & .0003 \\
  rs596134\_G & -.0090(.0202) & -.0094(.205) & .0004 &  -.0094(.0203) & .0004 \\
  rs3741311\_G & .0316(.0215) & .0311(.0364) & .0005 & .0298(.0216) & .0018 \\
  rs1941637\_A & -.0158(.054) & -.0159(.0214) & .0001 & -.0147(.0542) & .0011 \\
  rs658624\_G & -.0134(.0205) & -.136(.0203) & .0002 & -.0131(.0206) & .0003 \\
  rs625464\_A & .0232(.0200) & .0234(.286) & .0002 & .0236(.0200) & .0004 \\
  rs11216788\_A & .0057(.0204) & .0057(.219) & .0000 & .0061(.0204) & .0004 \\
  rs679327\_G & -.0094(.0217) & -.0092(.251) & .0002 & -.0092(.0218) & .0002 \\
  rs678957\_A & .0061(.0457) & .0053(.0209) & .0008 & .0066(.0459) & .0005 \\
  rs7944321\_A & -.0238(.0252) & -.0238(.0327) & .0000 & -.0230(.0253) & .0008 \\
  rs7949751\_G & .0258(.0233) & .0257(.0267) & .0001 & .0254(.0234) & .0004 \\
  rs11216816\_A & -.0406(.0268) & -.0399(.0201) & .0007 & -.0392(.0268) & .0014 \\
  rs1805\_A & -.0061(.0200) & -.0060(.0201) & .0001 & -.0056(.0201) & .0005 \\
 rs3759001\_A & -.0101(.0205) & -.103(.214) & .0002 & -.0098(.0206) & .0003 \\
rs4938493\_C & .0044(.0203) & .0044(.217) & .0000 & .0039(.0204) & .0005 \\
 rs2853009\_A & .0068(.0201) & .0068(.0222) & .0000 & .0064(.0202) & .0004 \\
 rs12282721\_A & -.0050(.0327) & -.0052(.0199) & .0002 & -.0072(.0328) & .0022 \\
 rs676134\_A & .0192(.0210) & .0185(.0200) & .0007 & .0173(.0210) & .0019 \\
 rs7083\_A & -.0570(.0205) & -.0572(.0205) & .0002 & -.0551(.0206) & .0019 \\
\hline
\end{tabular}
}
\end{center}
\end{table}

\begin{table}[h]
\begin{center}
\caption{Leiden Longevity Study: Estimates of the association between the 41 selected SNPS and triglyceride levels for men and for three different approaches. The retrospective likelihood approach with same variance assumed for the shared random effect, with different variances, and the naive approach. Are also presented the absolute difference between the estimates of the two last approaches with the first one. Into brackets are the standard errors.}
\scalebox{0.7}{
\begin{tabular}{rccccc}
\hline
SNPs & Constrained approach & \multicolumn{2}{c}{Not Constrained approach} & \multicolumn{2}{c}{Naive Likelihood}  \\
 & Estimates (SE) & Estimates (SE) & Difference & Estimates (SE)& Difference\\
\hline
rs3863318\_A & .0511(.0236) & .0501(.0236) & .0010 & .0494(.0236) & .0017 \\
  rs2512139\_A & .0294(.0249) & .0285(.0249) & .0009 & .0294(.0249) &.0000 \\
  rs7103514\_G & -.0329(.0426) & -.0324(.0427) & .0005 & -.0332(.0426) & .0003 \\
  rs2512158\_A & .0119(.0241) & .0118(.0241) & .0001 & .0116(.0241) & .0003 \\
  rs11216648\_G & -.0121(.0244) & -.0124(.0244) & .0003 & -.0114(.0244) & .0007 \\
  rs4936414\_G & .0161(.0244) & .0162(.0245) & .0001 & .0161(.0244) & .0000 \\
  rs4252287\_A & .0624(.0386) & .0631(.0386) & .0007 & .0620(.0386) & .0004 \\
  rs947889\_G & .0343(.0237) & .0335(.0238) &  .0008 & .0335(.0237) & .0008 \\
  rs2512154\_A & .0458(.0351) & .0444(.0353) & .0014 & .0453(.0352) & .0009 \\
  rs4936417\_G & .0280(.0256) & .0280(.0256) & .0000 & .0283(.0256) & .0003 \\
  rs652107\_G & -.0562(.0310) & -.0568(.0311) & .0006 & -.0579(.0310) & .0017 \\
  rs12576767\_A & -.0087(.0468) & -.0086(.0472) & .0001 & -.0080(.0469) & .0007 \\
  rs1786186\_A & -.0755(.0356) & -.0748(.0357) & .0007 & -.0736(.0357) & .0019 \\
  rs3825050\_A & .0178(.0250) & .0182(.0252) & .0004 & .0177(.0251) & .0001 \\
  rs10160375\_G & .0172(.0248) & .0172(.0249) & .0000 & .0170(.0248) & .0002 \\
  rs689264\_A & -.0549(.0466) & -.0549(.0465) & .0000 & -.0536(.0466) & .0013 \\
  rs948461\_A & .0360(.0249) & .0352(.025) & .0008 & .0360(.0249) & .0000 \\
  rs2276123\_A & .0121(.0253) & .0121(.0253) & .0000 & .0121(.0253) & .0000 \\
  rs948466\_A & -.0493(.0498) & -.0491(.0499) & .0002 & -.0470(.0499) & .0023 \\
  rs881122\_A & -.0519(.0280) & -.0523(.028) & .0004 & -.0526(.0280) & .0007 \\
  rs2276129\_A & .0161(.0250) & .0161(.0251) & .0000 & .0162(.0250) & .0001 \\
  rs2155857\_A & .0473(.0633) & .0482(.0634) & .0009 & .0526(.0633) & .0053 \\
  rs1894177\_A & .0267(.0232) & .0263(.0233) & .0004 & .0273(.0232) & .0006 \\
  rs596134\_G & .0164(.0238) & .0160(.0238) & .0004 & .0172(.0238) & .0008 \\
  rs3741311\_G & -.0054(.0247) & -.0052(.0248) & .0002 & -.0046(.0247) & .0008 \\
  rs1941637\_A & .0888(.0667) & .0897(.0668) & .0011 & .0921(.0667) & .0033 \\
  rs658624\_G & .0235(.0239) & .0232(.0239) & .0003 & .0247(.0239) & .0012 \\
  rs625464\_A & -.0245(.0239) & -.0241(.024) & .0004 & -.0249(.0240) & .0004 \\
  rs11216788\_A & .0575(.0239) & .0572(.024) & .0003 & .0579(.0240) & .0004 \\
  rs679327\_G & -.0622(.0270) & -.0623(.0271) & .0001 & -.0621(.0270) & .0001 \\
  rs678957\_A & -.0756(.0583) & -.0768(.0585) &  .0012 & -.0761(.0583) & .0005 \\
  rs7944321\_A & -.1010(.0315) & -.1012(.0315) & 0002 & -.1008(.0315) & .0002 \\
  rs7949751\_G & -.0311(.0270) & -.0310(.027) & .0001 & -.0303(.0270) & .0008 \\
  rs11216816\_A & -.0334(.0330) & -.0333(.033) & .0001 & -.0331(.0330) & .0003 \\
  rs1805\_A & -.0430(.0243) & -.0428(.0243) & .0002 & -.0427(.0243) & .0003 \\
  rs3759001\_A & -.0408(.0241) & -.0406(.0242) & .0002 & -.0407(.0241) & .0001 \\
  rs4938493\_C & .0377(.0236) & .0377(.0237) & .0000 & .0379(.0236) & .0002 \\
  rs2853009\_A & .0184(.0234) & .0184(.0234) & .0000 & .0187(.0234) & .0003 \\
  rs12282721\_A & -.0532(.0379) & -.0534(.038) & .0002 & -.0523(.0379) & .0009 \\
  rs676134\_A & -.0236(.0243) & -.0234(.0244) & .0002 & -.0233(.0243) & .0003 \\
  rs7083\_A & .0228(.0244) & .0229(.0245) & .0001 & .0221(.0244) & .0007 \\
  \hline
\end{tabular}
}
\end{center}
\end{table}

\begin{figure}[h]
\centering\includegraphics[scale=0.45]{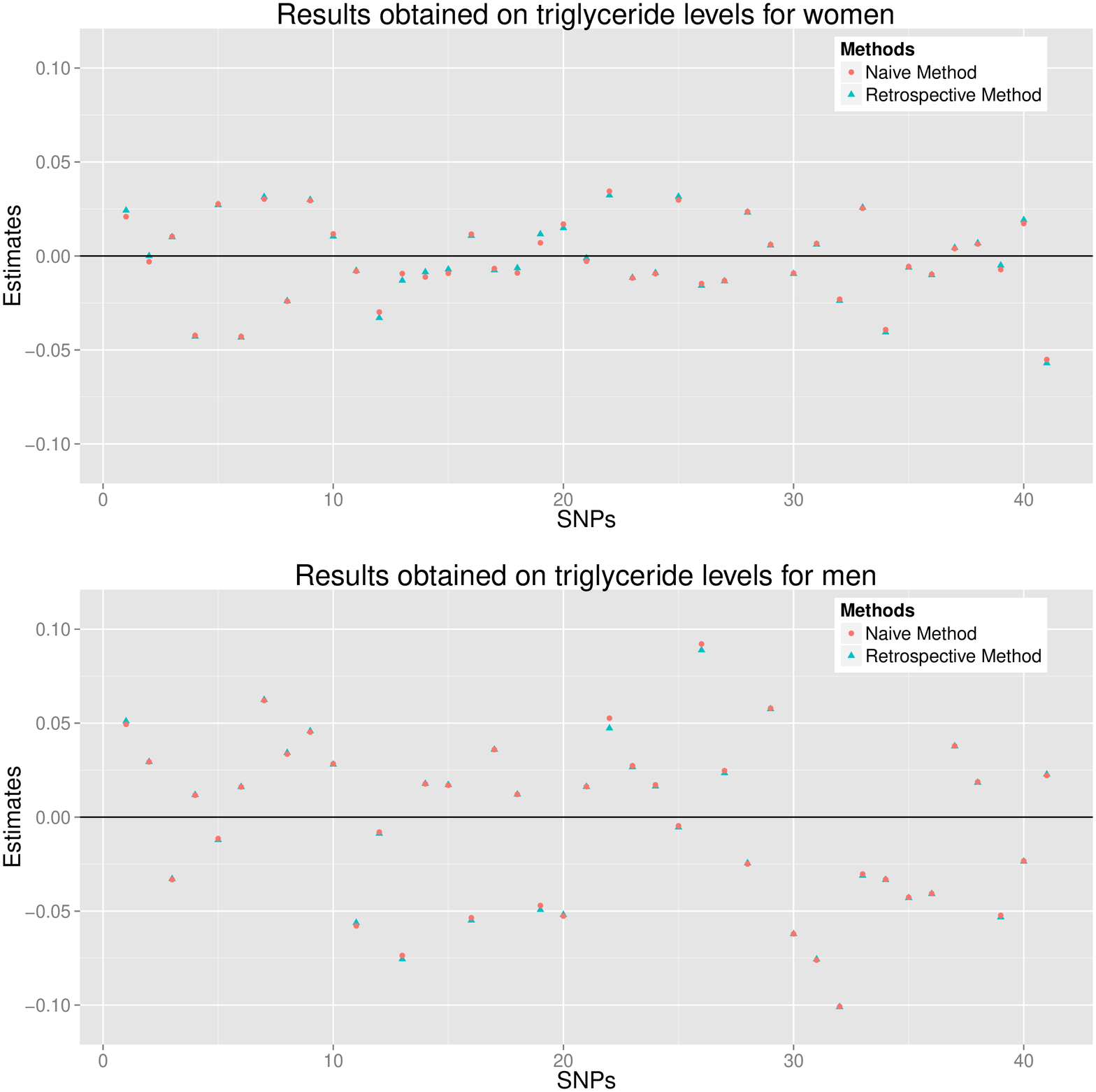}
\caption{Estimates for the association of 41 SNPs with triglyceride in the LLS. In the top and bottom are the estimates of the 41 SNPs  for women and for men respectively. The black line represents no SNP effect on triglyceride.}
\label{SNPS}
\end{figure}

\end{document}